\documentclass[smallextended]{svjour3}       
\usepackage{graphicx}
\usepackage{amsmath,amssymb}

\newcommand{\bea}{\begin{eqnarray}}
\newcommand{\eea}{\end{eqnarray}}
\newcommand{\be}{\begin{equation}}
\newcommand{\ee}{\end{equation}}

\usepackage{color}

\begin{document}

\title{Status of the asymptotic safety paradigm for quantum gravity and matter
}

\author{Astrid Eichhorn 
}

\institute{Institute for Theoretical Phyiscs, University of Heidelberg \\
Philosophenweg 16, 69120 Heidelberg, Germany\\
              \email{a.eichhorn@thphys.uni-heidelberg.de}           
}

\date{Received: date / Accepted: date}

\maketitle

\begin{abstract}
In the asymptotic safety paradigm, a quantum field theory reaches a regime with quantum scale invariance in the ultraviolet, which is described by an interacting fixed point of the Renormalization Group. Compelling hints for the viability of asymptotic safety in quantum gravity exist, mainly obtained from applications of the functional Renormalization Group. The impact of asymptotically safe quantum fluctuations of gravity at and beyond the Planck scale could at the same time induce an ultraviolet completion for the Standard Model of particle physics with high predictive power.
\end{abstract}

\section{Building blocks of the universe: Beyond effective field theory}
\label{sec:intro}
What are the microscopic building blocks of the universe? Our most successful models of the building blocks of matter and their interactions -- General Relativity (GR) and the Standard Model (SM) of particle physics -- are powerful effective theories that only hold over a finite range of scales and thus do not answer this question. In GR,  the existence of singularities in physical solutions signals a breakdown, and quantum gravity is expected to provide a resolution. Due to the dimensionful nature of Newton's coupling, $[G_N] = 2-d$ in $d$ spacetime dimensions, the perturbative quantization of GR appears to require the introduction of new counterterms at every loop order \cite{'tHooft:1974bx}. The couplings of the counterterms are free parameters that must be fixed by experiment. At infinitely high loop infinitely many free parameters imply a breakdown of predictivity.
Nevertheless, one can derive predictions of quantum gravity, such as the quantum correction to the gravitational potential between two point sources  \cite{Donoghue:1993eb}, at
 energies $E$ below the Planck mass $M_{\rm Pl}$ where the contribution of higher-order interactions is generically suppressed by positive powers of $(E/M_{\rm Pl})$. Predictivity is lost at $E\approx M_{\rm Pl}$, where all higher-order terms are expected to contribute equally. Thus, a predictive model of quantum gravity exists below the Planck scale, and misses an ultraviolet (UV) completion.

In contrast to Einstein-Hilbert gravity, the Standard Model is perturbatively renormalizable, i.e., the finitely many counterterms are of the same form as the terms in the original action. However, perturbative renormalizability is actually neither necessary nor sufficient to obtain a fundamental theory, i.e., one that holds up to arbitrarily large energy scales. The simplest example is a $\lambda_4/8\, \phi^4$ model in $d=4$: Absorbing the one-loop divergence in a renormalization of $\lambda_4$ leads to a Renormalization Group (RG) scale dependence given by the beta function encoding the dimensionless scale-derivative of the coupling
\be
\beta_{\lambda_4} = \frac{3}{16\pi^2}\lambda_4^2+...\, \, .
\ee
This implies a logarithmically increasing running coupling, resulting in a divergence at a finite momentum scale, the Landau pole, which signals the breakdown of perturbation theory. Beyond perturbation theory, the conclusion that the theory cannot be extended beyond the scale of the Landau pole appears to persist, and is known as the triviality problem: Demanding that the coupling is finite over
 an infinite range of scales results in a vanishing coupling in the infrared (IR), i.e., a trivial theory \cite{Frohlich:1982tw}. Perturbatively renormalizable models that are fundamental such as, e.g., Yang-Mills theories in $d=4$, require that the running coupling is asymptotically free, i.e. vanishes in the far UV.

The triviality problem most likely carries over to the Higgs-Yukawa and Abelian hypercharge sector of the Standard Model \cite{triviality},\cite{GellMann:1954fq}. The corresponding Landau poles, which appear to signal the need for new physics, lie beyond the Planck scale. Thus, quantum gravity might be the new physics required to solve the triviality problem and to render the Standard Model UV complete. 

\section{Asymptotic safety: The main idea}
\label{label:secAS}
Asymptotic safety generalizes asymptotic freedom and both can underlie predictive fundamental QFTs. The latter implies that a model reaches a regime where it becomes free in the UV, and hence protected from the triviality problem.
The free model is a Renormalization Group (RG) fixed point: When all interactions vanish at some energy scale, they vanish everywhere. Asymptotic safety generalizes this to an RG fixed point at finite coupling, i.e., residual interactions exist. The fixed point is reached in the dimensionless versions of all couplings $g_i$. For a coupling $\bar{g}_i$ with canonical mass dimension $d_{\bar{g}_i}$ its dimensionless counterpart is $g_i= \bar{g}_i\, k^{-d_{\bar{g}_i}}$. If all $g_i$ reach an asymptotically safe fixed point, this implies finiteness of observables \cite{Weinberg:1980gg}.
\begin{figure*}
\begin{center}
\includegraphics[width=0.4\linewidth]{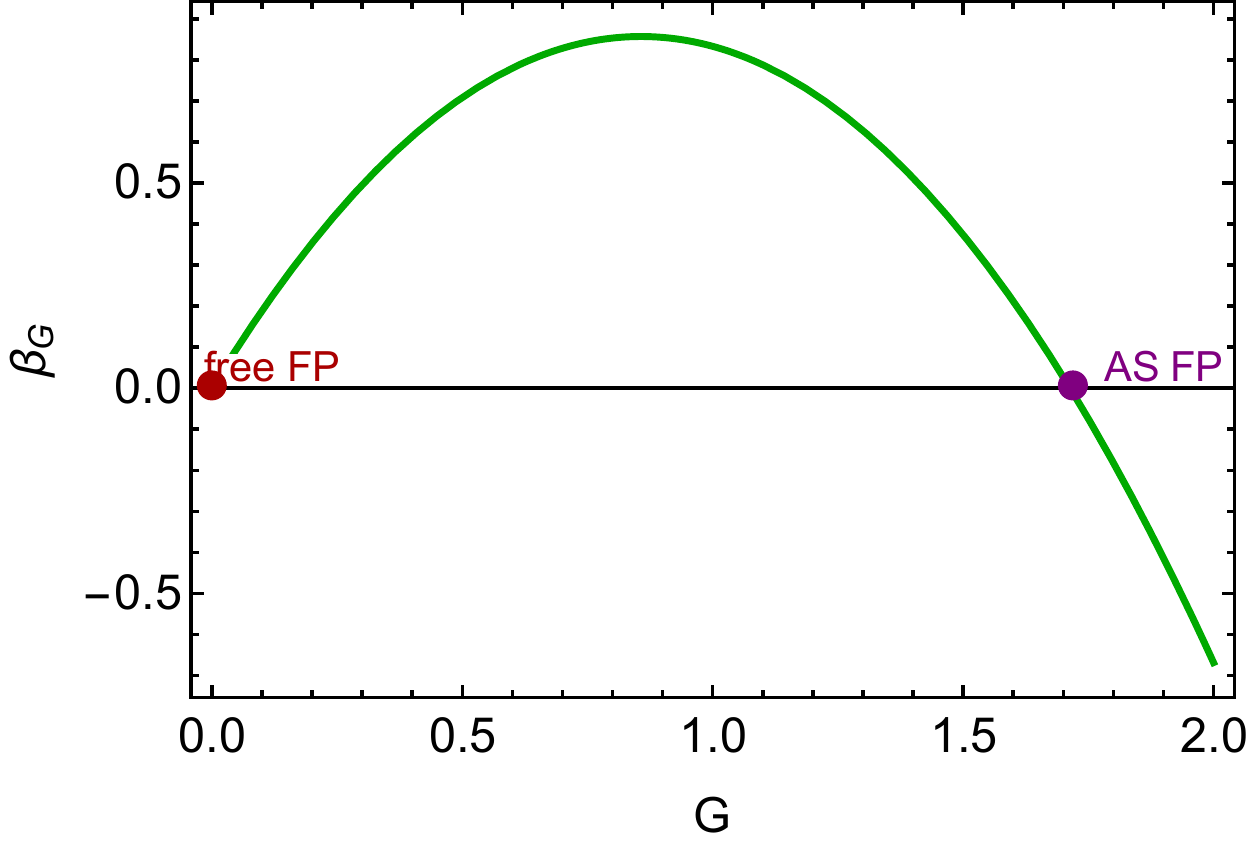} \quad \includegraphics[width=0.4\linewidth]{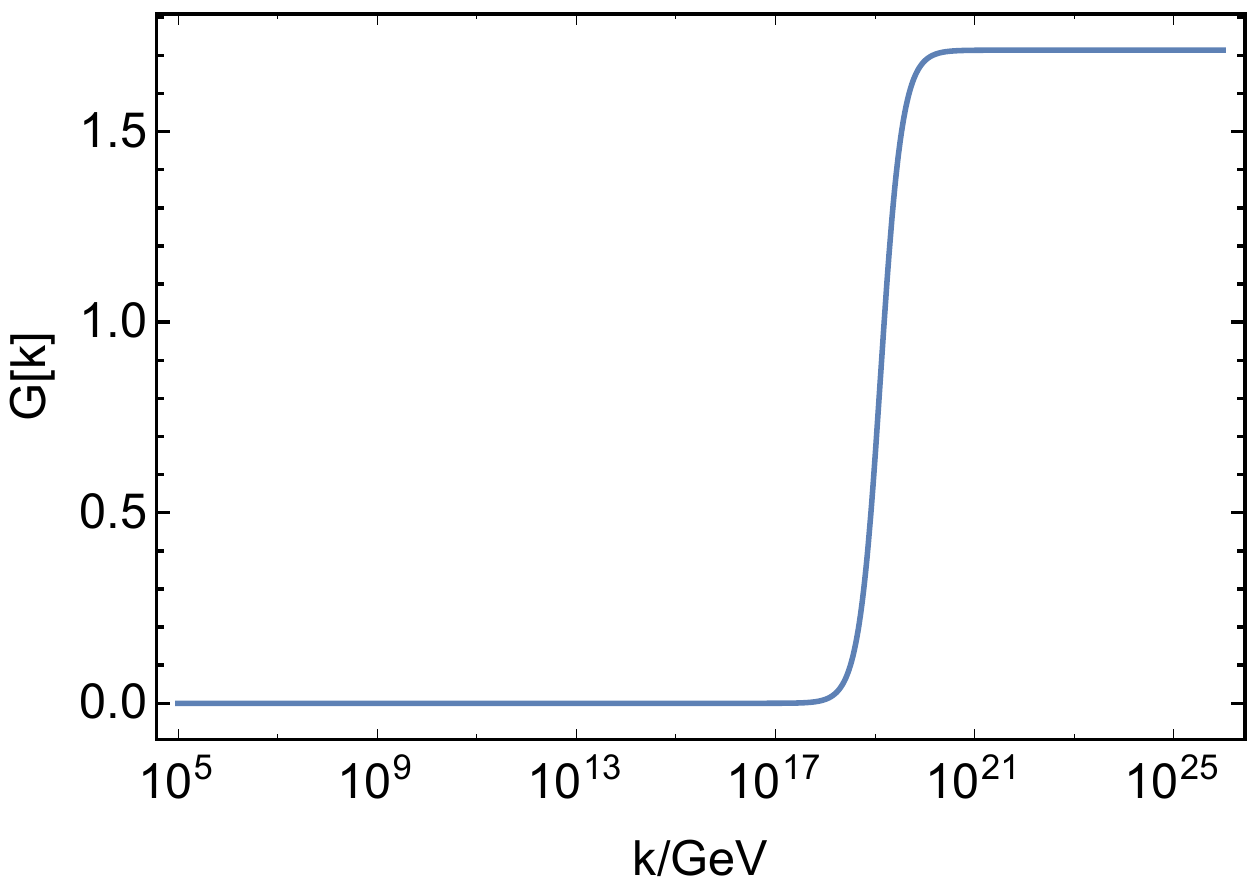}\\
\includegraphics[width=0.4\linewidth]{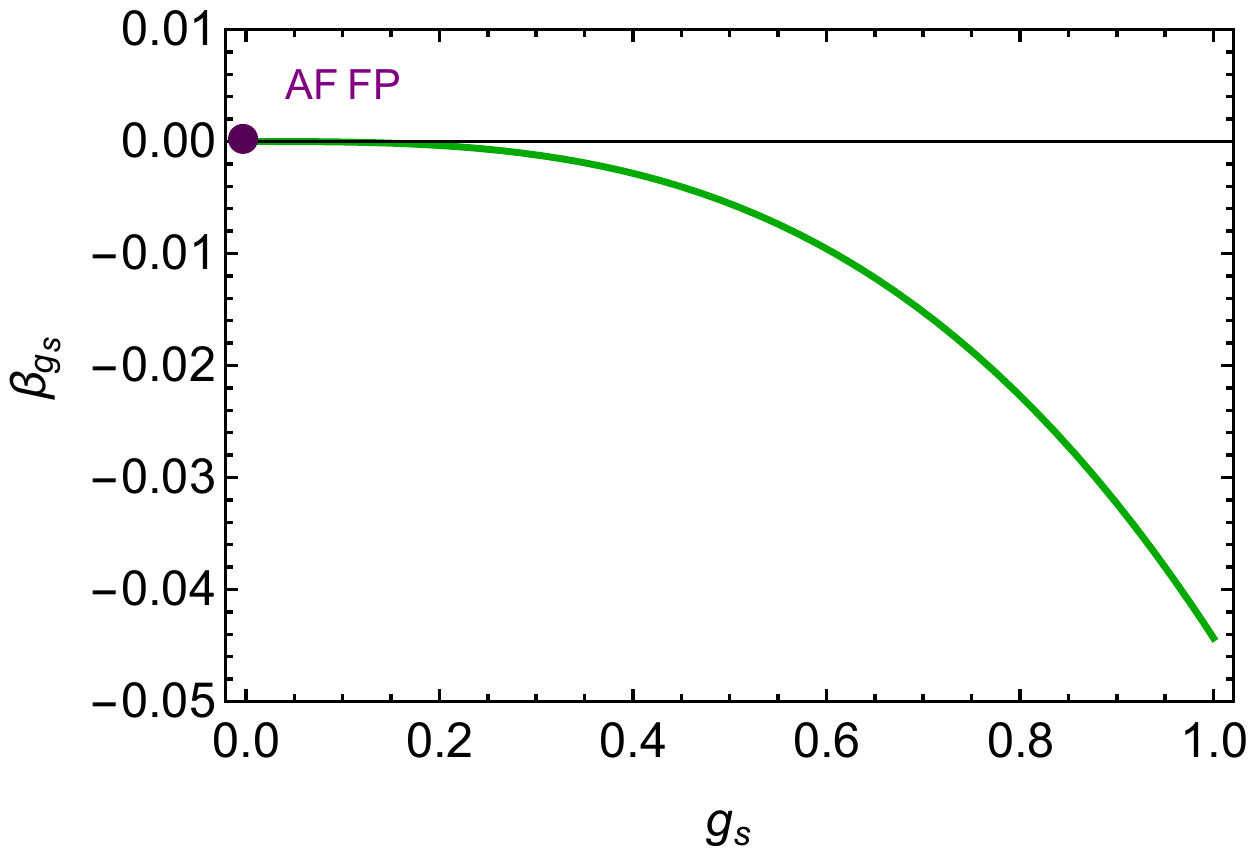}\quad \includegraphics[width=0.4\linewidth]{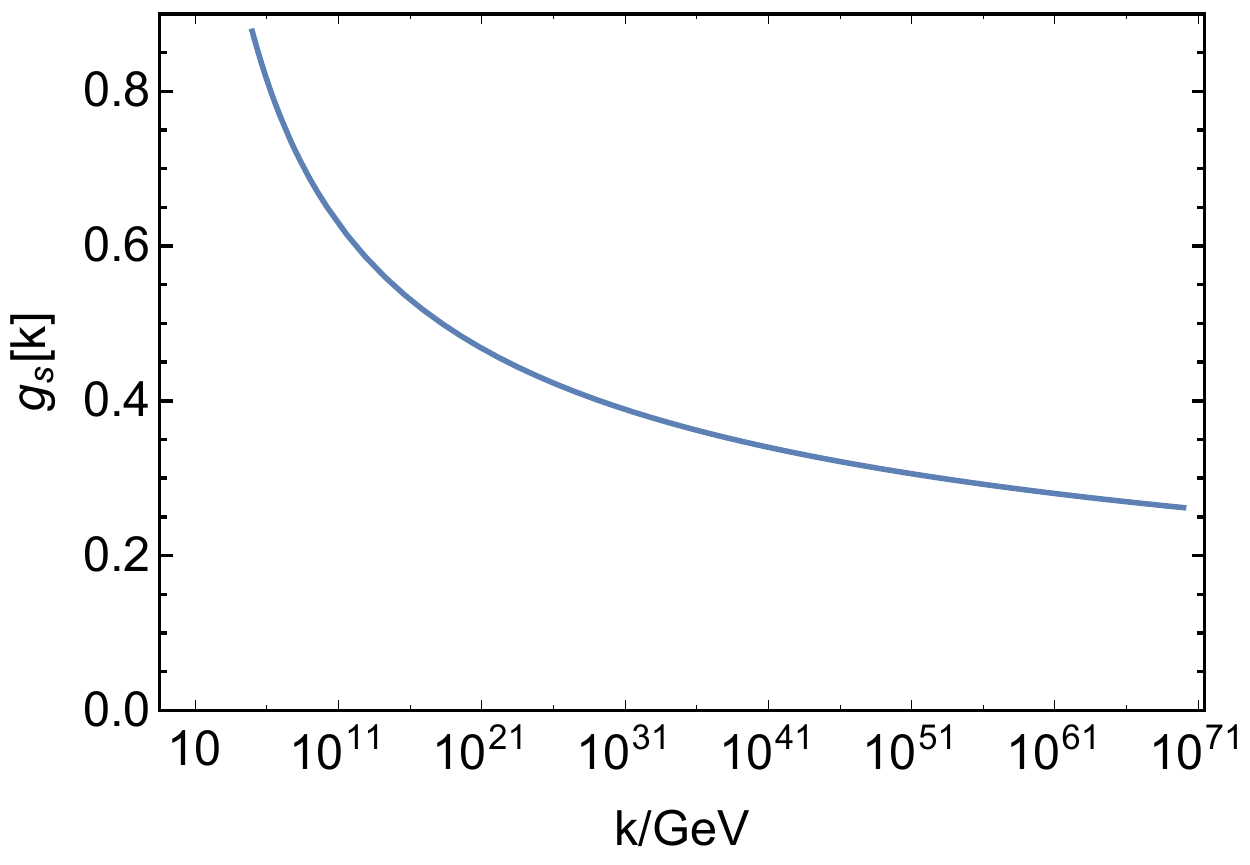}
\caption{\label{fig:betas_and_coups} Beta functions and the corresponding running for a UV attractive free/interacting fixed point (lower/upper panels). As the fixed points are UV attractive, the value of the coupling at one reference scale is an initial condition that can be chosen freely. Upper panels: beta function and scale dependence of the Newton coupling according to Eq.~\eqref{eq:glambdaflow}.}
\end{center}
\end{figure*}
As fixed points of the RG flow, asymptotic safety/freedom imply quantum scale invariance, cf.~Fig.~\ref{fig:betas_and_coups}. Such regimes can be attained in models with dimensionfull couplings, since the scaling of couplings is determined by the sum of canonical scaling and quantum scaling which is a consequence of loop effects.
 In fact, a fixed point can arise from a balance between canonical and quantum scaling, as 
 \be
\beta_{g_i}= -d_{\bar{g}_i}\, g_i + \eta_i (g_i),\label{eq:betaschem}
\ee
where $\eta_i$ is an anomalous scaling dimension that arises as a consequence of quantum fluctuations. An interacting fixed point lies at
\be
g_i^{\ast} =  \eta_i/d_{\bar{g}_i}.
\ee 
For instance, in $d=4-\epsilon$, the $\lambda_4\phi^4$ theory has a fixed point at $\lambda_4^{\ast}=16\pi^2 \epsilon/3$, the Wilson-Fisher fixed point \cite{Wilson:1971dc} playing an important role in statistical physics. 
In that setting, interacting fixed points encode the scaling exponents near a continuous (second or higher order) phase transition, where scale invariance is due to a diverging correlation length at criticality.

\begin{figure}
\begin{minipage}{0.5\linewidth}
\includegraphics[width=0.8\linewidth, clip=true, trim=11cm 5cm 13cm 13cm]{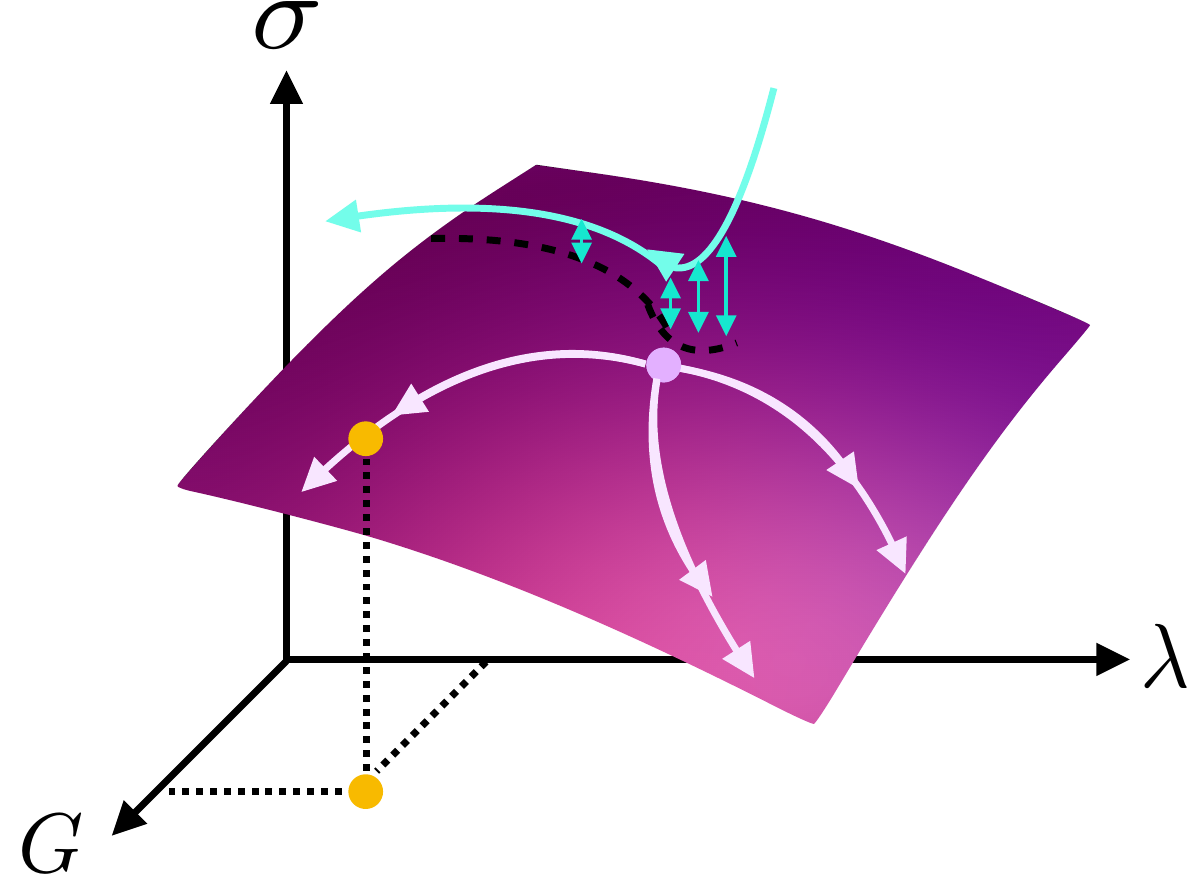}
\end{minipage}\hfill
\begin{minipage}{0.45\linewidth}
\caption{\label{fig:crithypersurface}Illustration of a 3-dimensional subspace of theory space, with trajectories emanating from an asymptotically safe fixed point. $\sigma$ is determined at all scales in terms of the $G, \lambda$, as the critical hypersurface is 2-dimensional. An RG trajectory that is displaced from the value of $\sigma$ determined by $G, \lambda$ is not UV safe.}
\end{minipage}
\end{figure}

QFTs live in theory space, which is the infinite-dimensional space of all couplings that are compatible with the symmetries of the model.  Asymptotic safety/freedom is the existence of a fixed point in this space, i.e., for a model to become asymptotically safe, \emph{all} couplings have to reach a scale-invariant fixed point.
Thus, determining whether a model can become asymptotically safe/free, requires us to explore the RG flow of all infinitely many couplings.\\
On the other hand, in perturbatively renormalizable models with asymptotic freedom one typically does not think about higher-order couplings, but restrict the setting to the perturbatively renormalizable ones, thus seemingly working in a finite-dimensional theory space. The reason is not the absence of higher-order couplings in a Wilsonian view of such models. However, the higher-order couplings are determined in terms of the perturbatively renormalizable ones, i.e., they are not free parameters of the model. 
This is an example of a model that reaches a fixed point and has a finite number of free parameters although the fixed point lies in the infinite dimensional theory space. The free parameters of the model -- which are the relevant, i.e., UV-attractive couplings, span the UV-critical hypersurface of the fixed point. A flow that emanates from the fixed point can only move within the UV-critical hypersurface, i.e., in the direction of the relevant couplings -- any displacement in a UV repulsive direction prevents the couplings from being able to continuously flow "back" into the fixed point towards the UV, cf.~Fig.~\ref{fig:crithypersurface}.
As the relevant couplings are the free parameters of the model, predictivity requires a finite-dimensional UV-critical hypersurface. All infinitely many directions which are orthogonal to the (local tangent space of the) hypersurface at the fixed point are predictable. These concepts are central for the connection of asymptotic safety to phenomenology.

\begin{figure*}
\begin{minipage}{0.48\linewidth}
\includegraphics[width=0.8\linewidth]{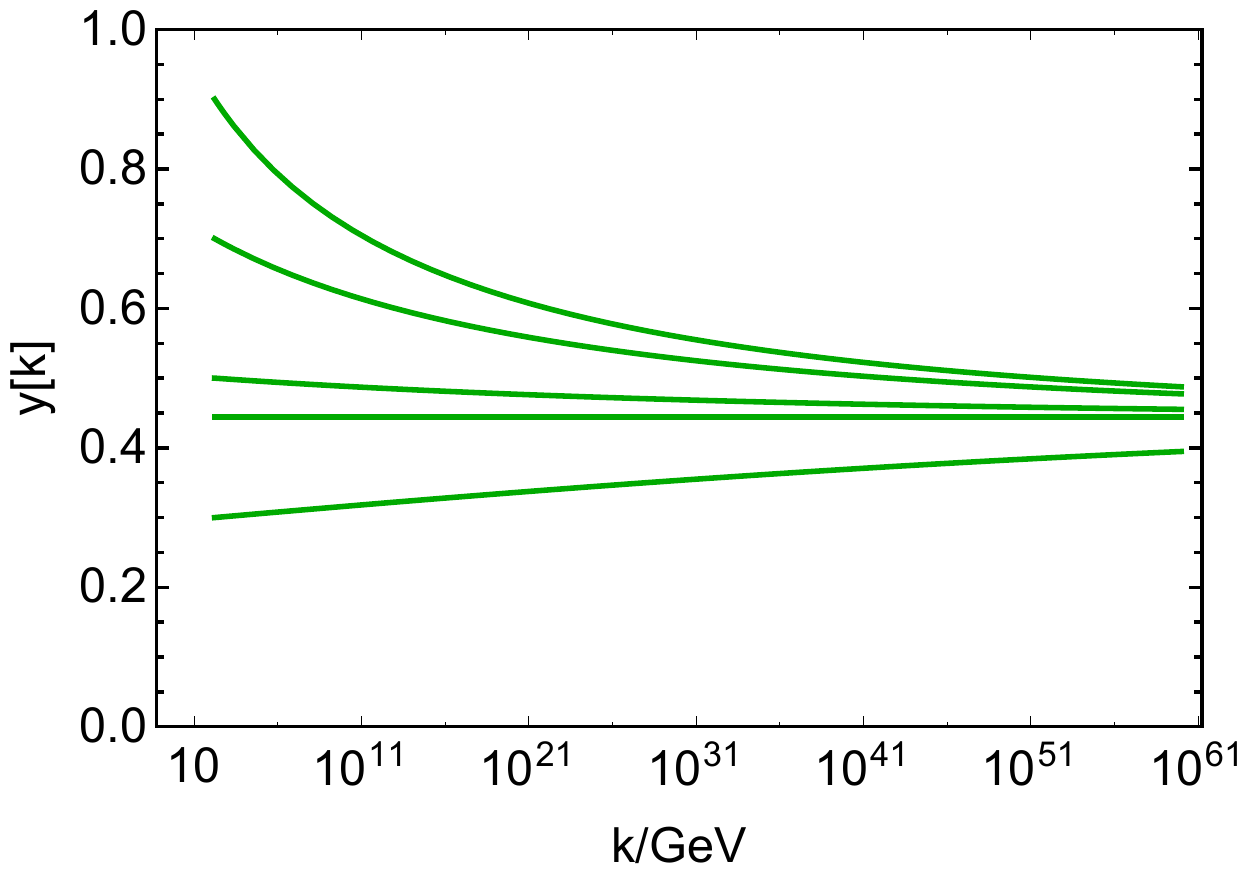} \\
 \includegraphics[width=0.8\linewidth]{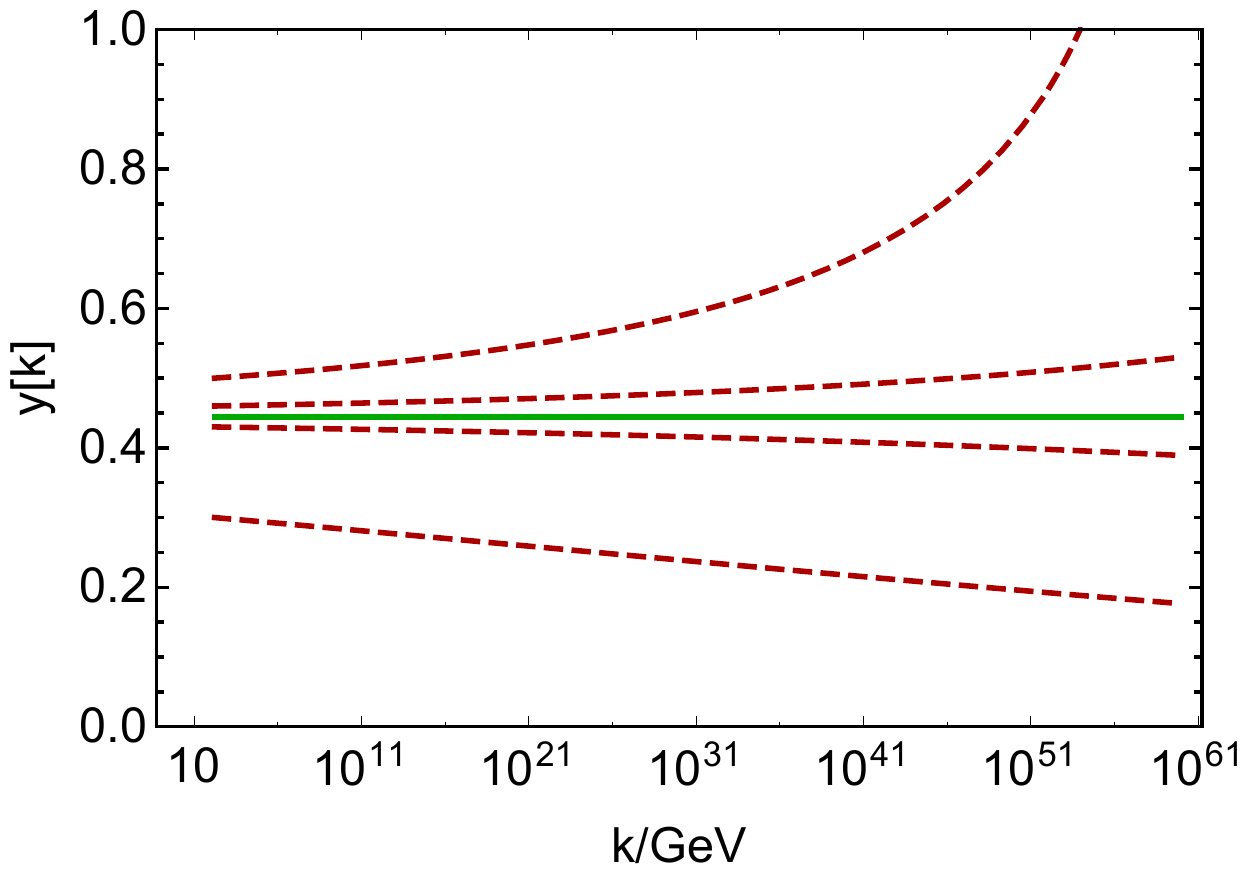}
\end{minipage}
\begin{minipage}{0.48\linewidth}
\includegraphics[width=\linewidth]{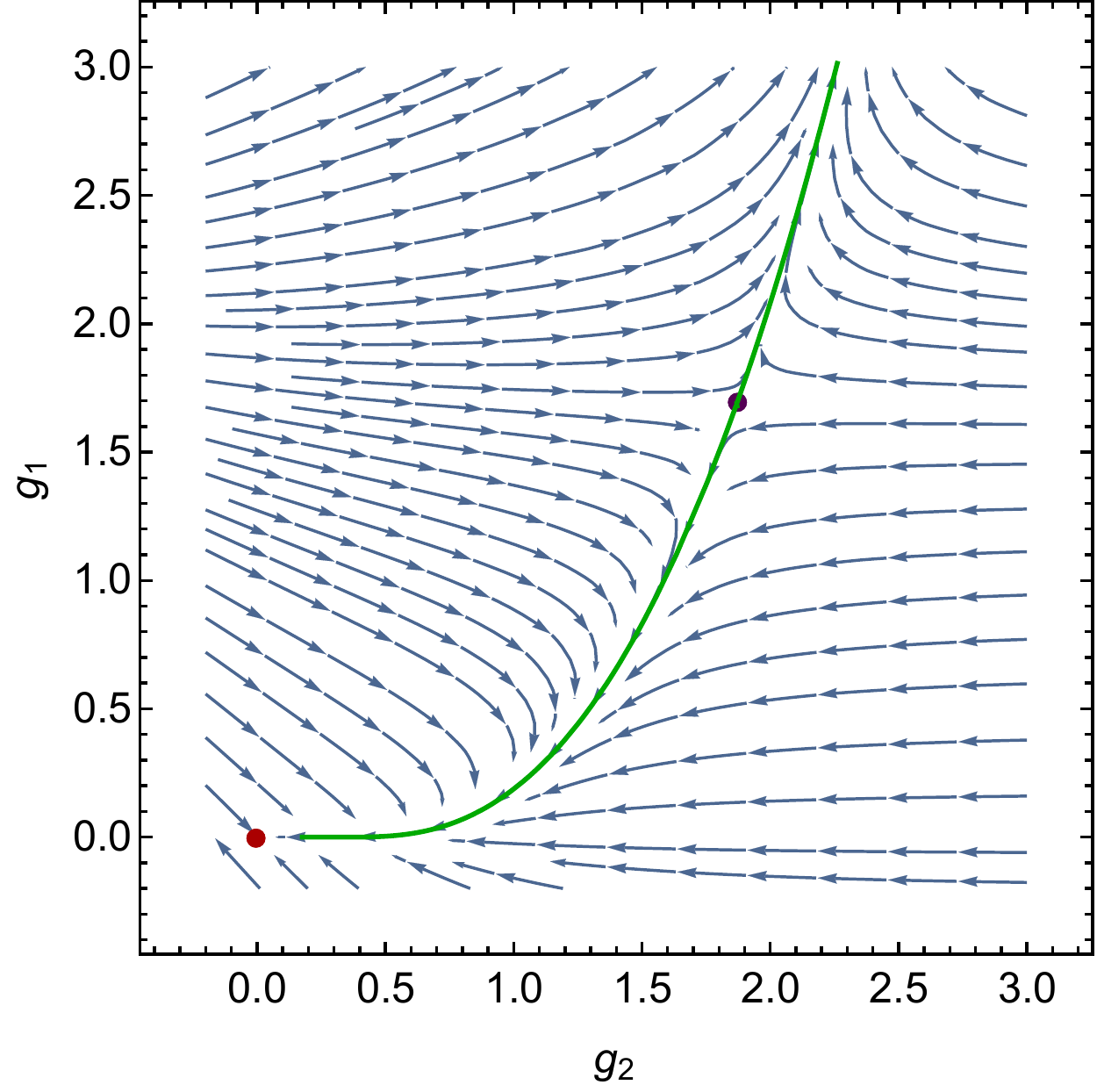}
\end{minipage}
\caption{\label{fig:irrelevant} Left upper(lower) panel: Illustration of an UV-attractive (repulsive) interacting fixed point: all (one unique) IR values are compatible with the fixed point. Right: Flow towards the IR with a fixed point (purple dot) at which $g_2$ is irrelevant: The critical hypersurface is one dimensional, i.e., the flow away from the fixed point is determined by the value of one coupling. The relevant direction is a superposition of $g_1$ and $g_2$, mostly aligned with $g_1$.}
\end{figure*}

At the fixed point, the relevant directions are IR-repulsive. Thus, RG trajectories emanating from the fixed point are parameterized by an increasing distance of the relevant couplings to their fixed-point values, i.e., the values of the relevant couplings encode the deviations from scale invariance. The set of \emph{all} RG trajectories emanating from the fixed point spans the UV critical surface. Any point in this hypersurface can be reached in the IR by an RG trajectory that emanates from the fixed point in the UV. Accordingly, IR values of relevant couplings are not predicted, cf.~left upper panel in Fig.~\ref{fig:irrelevant}, and have to be determined by an experiment. 
In contrast, all directions orthogonal to the critical hypersurface -- the irrelevant ones --  are IR attractive. Thus, a tiny deviation from the fixed point to a point \emph{outside} the critical surface would lead to a flow approaching the critical surface to the IR. Therefore, there is no additional freedom in the IR values of the irrelevant couplings for a UV complete model. Note however that in one picks an RG trajectory that is perturbed away from the critical hypersurface by some (tiny) amount at some scale $k$, this trajectory \emph{cannot} be reached from the fixed point: As the fixed point is UV repulsive in the irrelevant couplings, reversing the flow and following it towards the UV leads to an \emph{increase} in the distance to the fixed point, cf.~left lower panel in Fig.~\ref{fig:irrelevant}. Thus, an RG trajectory that does not lie exactly within the critical hypersurface cannot be UV complete.
 Hence, the IR values of irrelevant couplings provide tests of the underlying microscopic model: Only one unique value of an irrelevant coupling is compatible with an underlying microscopic fixed point, cf.~lower left panel in Fig.~\ref{fig:irrelevant}. For instance, in Fig.~\ref{fig:crithypersurface} the value of $\sigma$ at any scale in determined by the values of $G, \lambda$ at that scale, i.e., all UV complete RG trajectories in this three-dimensional space are fully characterized in terms of two parameters
 As the critical hypersurface can be curved, an irrelevant coupling does not need to stay constant, i.e., curvature of the critical hypersurface forces an irrelevant coupling to run as a function of scale. The key point is that its value \emph{at all scales} is determined by  the relevant couplings at the corresponding scale. Note that (ir)relevant directions need not be aligned with the couplings in the action, i.e., superpositions can be (ir)relevant. At a free fixed point, the irrelevant directions have negative mass dimensionality, i.e., they are perturbatively nonrenormalizable. For an asymptotically free model, their values cannot be chosen freely, but are fixed by the relevant couplings, as discussed above. This does not mean that the couplings vanish in the IR, e.g., in QCD indications for gluon condensation are tied to a nontrivial ``potential" for the field-strength, with finite values of the couplings of higher-order operators \cite{gluecon}. The values of those couplings are determined in terms of \emph{one} free parameter, which is essentially the QCD scale.\\
Intuitively, a fixed point leads to predictions for the irrelevant couplings quantum fluctuations force the irrelevant coupling to stay at its fixed- point value all the way to the IR. For gravity coupled to matter, this will become particularly interesting: Quantum fluctuations of gravity force irrelevant matter couplings to remain glued to the critical hypersurface of the asymptotically safe gravity-matter fixed point, until quantum gravity is switched off dynamically, cf.~upper panel in Fig.~\ref{fig:betas_and_coups}. As this happens near the Planck scale,  the Planck-scale values of these couplings are fixed. Assuming no new physics between the Planck scale and the electroweak scale, one can predict IR physics from fixed-point physics, see, e.g., \cite{Shaposhnikov:2009pv,Harst:2011zx,Eichhorn:2017ylw,Versteegen}.

To determine which couplings are relevant, one linearizes the flow around the fixed point for which $\beta_{g_i}=0, \, \forall i$. The beta function for $g_i$ can depend on all other couplings $g_j$, summarized in the vector $\vec{g}$, and thus
\bea
\beta_{g_i} &=& \beta_{g_i}\Big|_{\vec{g}= \vec{g}^{\ast}}+ \sum_j\left(\frac{\partial \beta_{g_i}}{\partial g_j}\right)\Big|_{\vec{g}= \vec{g}^\ast} \left(g_j - g_{j}^{\ast} \right)+\mathcal{O}\left( \left(g_j - g_{j}^{\ast} \right)^2\right)\\
&=&0 + \sum_j \mathcal{M}_{ij}\left(g_j - g_{j}^{\ast} \right)+\mathcal{O}\left( \left(g_j - g_{j}^{\ast} \right)^2\right),\label{eq:stabm}
\eea
where the stability matrix $\mathcal{M}$ does not need to be diagonal. Thus one introduces coordinates $\hat{g}_i$ in theory space that are centered around the fixed point and in which $\mathcal{M}$ is diagonal with $-\theta_i$ as eigenvalues and $V^i$ as eigenvectors expressed in the $\vec{g}$ coordinates. Then, the linearized flow equation and its solution read
\bea
\partial_t \hat{g}_i&=& k\, \partial_k \hat{g}_i=- \theta_i \hat{g}_i,\\
\hat{g}_i(k) &=&  c_i\, \left(\frac{k}{k_0} \right)^{-\theta_i}, \mbox{ or }\quad g_i(k) = g_i^{\ast} + \sum_j c_j\, V_i^j\,  \left(\frac{k}{k_0} \right)^{-\theta_j}.\label{eq:linflow_hat}
\eea
Here, $c_i$ is a constant of integration and $k_0$ a reference scale. 
$\theta_i>0$ is associated to a relevant, i.e., UV-attractive direction. Then $c_i$ is a free parameter, which must be fixed by inferring the value of $\hat{g}_i$ from an experiment in the IR.
On the other hand, $\theta_i<0$ is associated to an irrelevant, i.e., UV-repulsive direction. To reach the fixed point in the UV, one has to demand $c_i=0$, thereby leaving no free parameter for $V^i$, i.e., for one linear combination of couplings.
For marginal directions with $\theta_i=0$, higher orders in Eq.~\eqref{eq:stabm} determine whether the direction is exactly marginal, or marginally (ir)relevant.

The critical exponents $\theta_i$ are linked to the canonical dimensionalities of couplings, as the beta functions contain a term  $\sim d_{\bar{g}_i}$ which enters the critical exponent, cf.~Eq.~\eqref{eq:betaschem}.
At the free fixed point, the critical exponents correspond exactly to the canonical dimensionality, thus couplings with (negative) positive mass dimensionality are (ir)relevant. Dimensionless couplings, such as, e.g., the gauge couplings in four dimensions, can become marginally irrelevant (as in QED, leading to the triviality problem) or marginally relevant (as in QCD, implying asymptotic freedom). At an interacting fixed point, the critical exponents receive additional contributions from residual interactions. A priori it is impossible to determine ir/relevance. However, one can expect interacting fixed points to be predictive, as there are only a few couplings with $d_{\bar{g}_i}\geq 0$, and all others have increasingly negative mass dimension. Thus, unless quantum contributions to the critical exponents grow with $i$, the canonical dimension ultimately "wins", yielding $\theta_i>0$ only for finitely many $i$.

\section{Functional Renormalization Group techniques}
\label{sec:FRG}
The functional Renormalization Group allows to probe the scale-dependence of a QFT, extract the beta functions and search for asymptotic safety. It is based on the definition of a scale-dependent effective dynamics encoded in an effective action $\Gamma_k$, that includes the effect of quantum fluctuations above the momentum scale $k^2$. 
$\Gamma_k$ contains all field monomials $\mathcal{O}_i$ respecting the underlying symmetries and contain positive integer powers of derivatives, multiplied by scale dependent couplings which span the theory space, 
\be
\Gamma_k = \sum_i \bar{g}_i(k)\mathcal{O}_i = \sum_i g_i\, k^{d_{\bar{g}_i}}\mathcal{O}_i.\label{eq:gamma_exp}
\ee
At a given scale $k$, the effective action defines a point in theory space. As $k$ is lowered to $k -\delta k$, quantum fluctuations with momenta in that range are included and lead to a change of $\Gamma_k$ along an RG trajectory. The beta functions can be extracted from  the scale derivative of the effective dynamics, as 
\be
\partial_t \Gamma_k =k \,\partial_k\Gamma_k= \sum_i \partial_t \bar{g}_i(k)\mathcal{O}_i = \sum_i \beta_{g_i} k^{d_{\bar{g}_i}}\mathcal{O}_i.
\ee
 Such an equation follows from the definition of $\Gamma_k$
 by a modified Legendre transform of the scale dependent generating functional $Z_k$
 \be
 Z_k[J] = \int_{\Lambda}\mathcal{D}\varphi\, e^{-S[\varphi] + \int_x J\, \varphi - \Delta S_k[\varphi]},\label{eq:Zkdef}
 \ee
  in terms of the expectation value $\phi= \langle \varphi \rangle$:
\bea
\Gamma_k [\phi]&=& {\rm extr}_J\left(\int_x J \, \phi - \ln Z_k\right)- \Delta S_k[\phi].\label{eq:defGammak}
\eea
The mass-like term $\Delta S_k$ depends on the regulator $R_k(p^2)$ and must suppress modes with $p^2<k^2$, i.e., only the high-energy modes contribute in Eq.~\eqref{eq:Zkdef},
\be
\Delta S_k[\varphi]=\frac{1}{2}\int \frac{d^dp}{(2 \pi)^d} \varphi(-p)R_k(p^2)\varphi(p).
\ee
Eq.~\eqref{eq:defGammak} implies the Wetterich equation for the scale-dependence of $\Gamma_k$ \cite{Wetterich:1992yh},
\be
\partial_t \Gamma_k:= k \partial_k \Gamma_k = \frac{1}{2}{\rm Tr} \left(\Gamma_k^{(2)}+R_k \right)^{-1}\partial_t R_k\quad\raisebox{-0.5cm}{\includegraphics[width=0.15\linewidth]{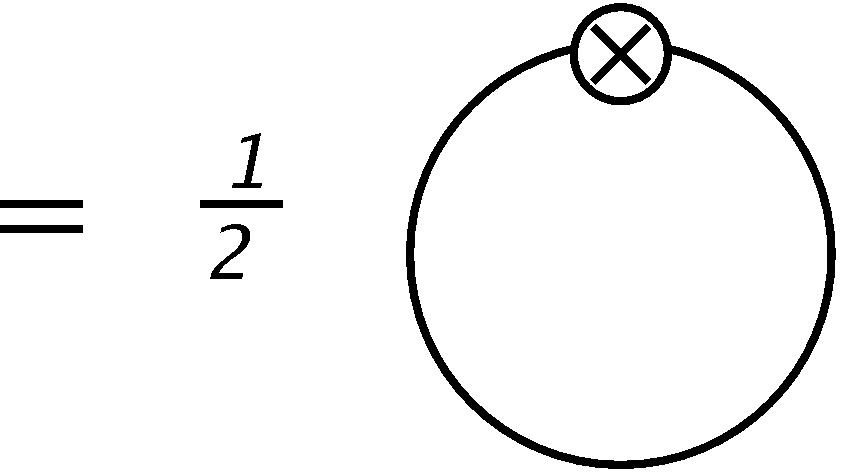}}\label{eq:Wetteq}
\ee
see also \cite{Morris:1993qb} and \cite{Berges:2000ew,Gies:2006wv} for reviews.
Herein $\Gamma_k^{(2)}$ is a shorthand for the second functional derivative of $\Gamma_k$ with respect to the field. The trace runs over the eigenvalues of the regularized propagator $(\Gamma_k^{(2)}+R_k)^{-1}$, i.e., for a simple scalar field on a flat background, it implements an integral over the loop momentum. \emph{Structurally} -- but not in the sense of perturbation theory -- eq.~\eqref{eq:Wetteq} has a diagrammatic representation as a one-loop equation with a loop over the full nonperturbative propagator with the regulator insertion.

\begin{figure*}
\includegraphics[width=0.45\linewidth]{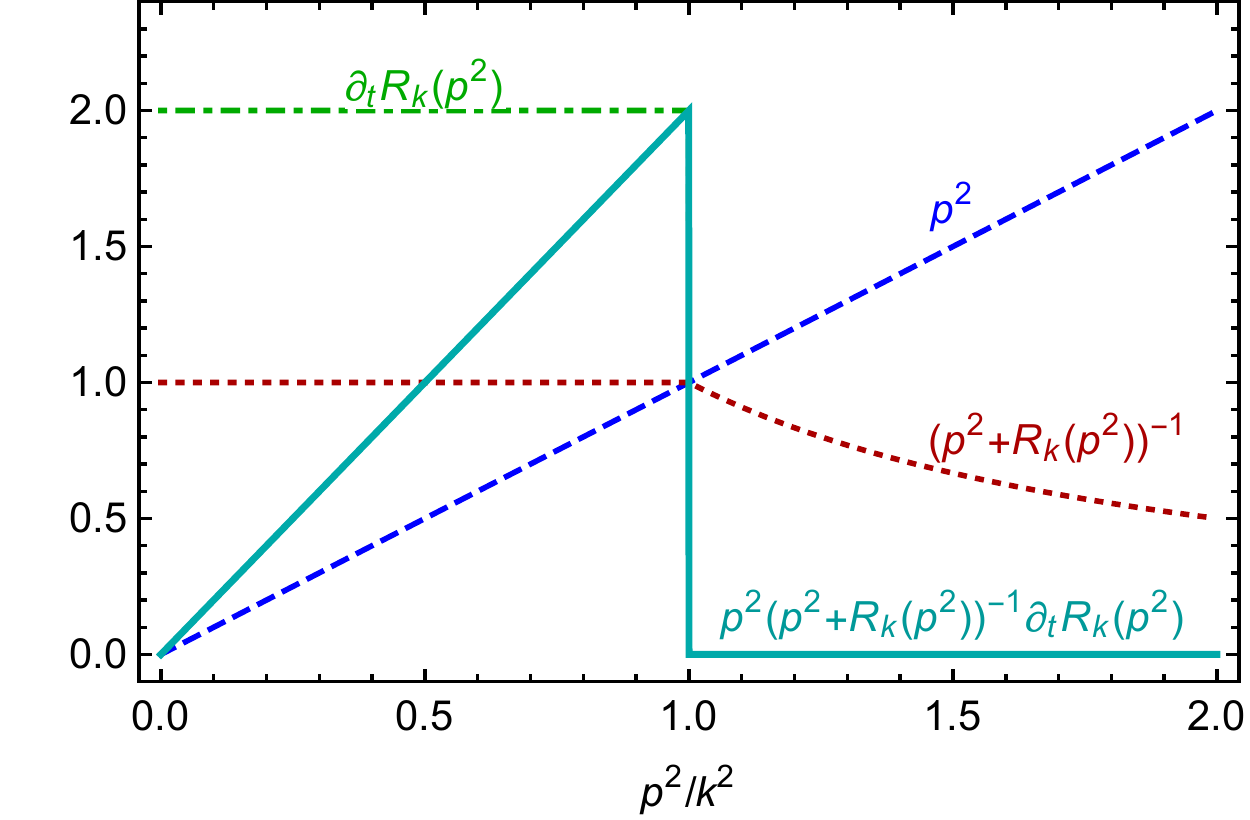}\quad \includegraphics[width=0.45\linewidth]{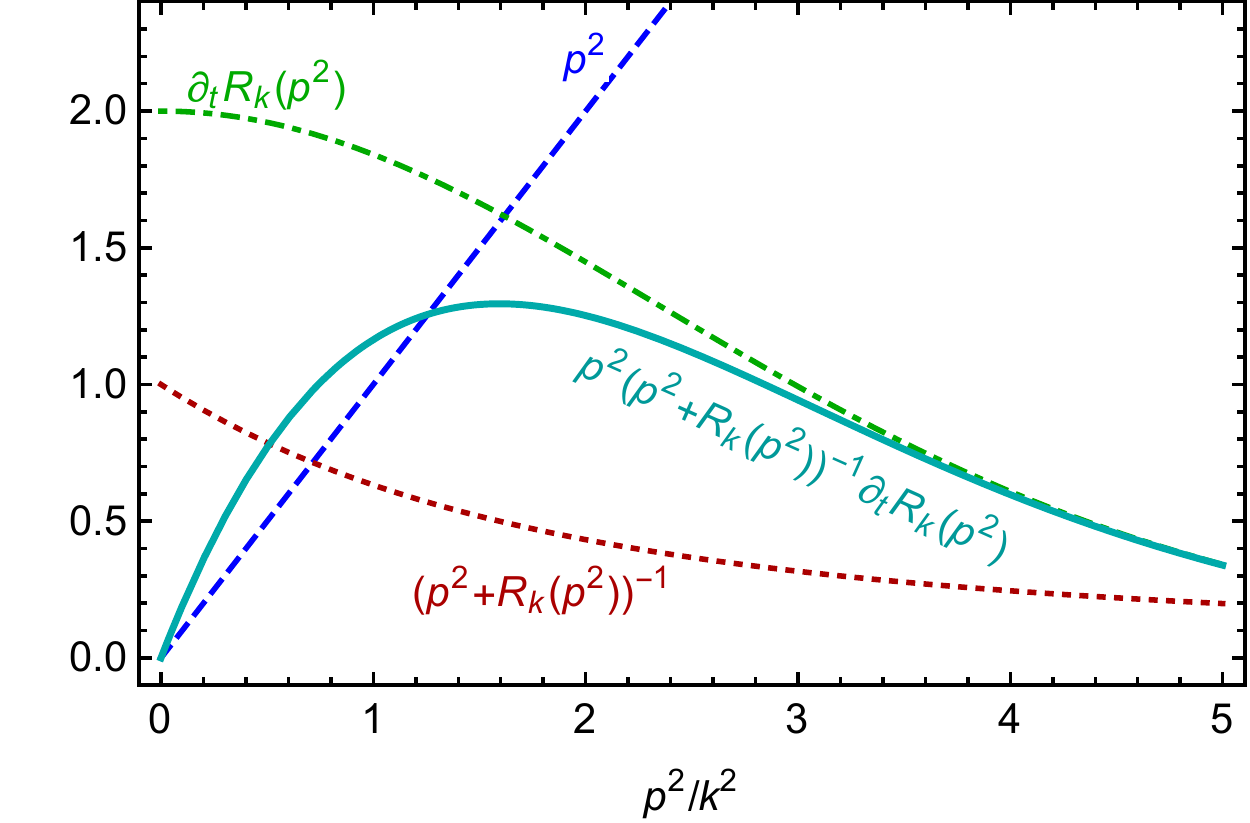}
\caption{\label{fig:regplots} On a flat background, $\rm Tr$ becomes a momentum integral with measure $p^2$ in $d=4$ (blue dashed). The full integrand (turquoise continuous) contains $\partial_t R_k$ (green dot-dashed) and $\Gamma_k^{(2)}+R_k$ (red dotted; $\Gamma_k^{(2)}=p^2$ in this example) has a peak at $p^2 \approx k^2$. The left panel shows the theta cutoff \cite{Litim:2001up}, $R_k=(k^2-p^2)\theta(k^2-p^2)$ and the right panel the exponential cutoff $R_k=p^2 /({\rm exp}(p^2/k^2)-1)$.}
\end{figure*}

The Wetterich equation is automatically UV and IR finite: The latter property is implemented by the mass-like regulator in the denominator, and the former follows as $\partial_t R_k$ vanishes for $p^2>k^2$. In fact, the trace over the eigenmodes of the propagator is peaked for modes with eigenvalues close to $k^2$, cf.~Fig.~\ref{fig:regplots}. Thus, the $k$-dependent change of the effective dynamics is driven by modes with momenta close to $k$, just as one would expect.

The Wetterich equation does not depend on the microscopic action $S$: It simply provides a vector field in theory space. Specifying a particular microscopic action provides an initial condition for the flow to $k=0$, yielding the corresponding effective action. However, the Wetterich equation is applicable more generally, allowing to search for viable microscopic dynamics, given a set of fields and symmetries. A consistent microscopic dynamics in a given theory space is  determined through the fixed point \cite{reconstruction}.

In practice one cannot work with the infinite sum in eq.~\ref{eq:gamma_exp}, and instead truncates, i.e.,  chooses a subspace of theory space that should contain the relevant operators. Accounting for deviations of the critical exponents from the canonical mass dimension, one expects a set with $d_{\bar{g}_i}\geq d_{\rm crit}$ to contain all relevant couplings. The value of $d_{\rm crit}$ can be tested in larger truncations by checking whether  couplings with $d_{\bar{g}_i}<d_{\bar{g}_i\,\rm crit}$ are in fact irrelevant. For instance, in gravity $d_{\bar{g}_i\,\rm crit}$ appears to be close to zero, cf.~Sec.~\ref{sec:ASstatusgrav}.

In truncations the Wetterich equation is not exact, i.e., operators beyond the truncation are generated on the right-hand-side, and can feed into the beta functions in the truncation. This contribution is set to zero in a truncation.
Thus one might question the reliability of the method. However, a vast collection of results on interacting fixed points in various QFTs (mostly in $d<4$) where comparisons with other techniques, e.g., $\epsilon$ expansion, Monte Carlo simulations and the conformal bootstrap are possible, highlights the power of the method.
Truncation-induced fixed points are usually unstable under extensions of the truncation. In contrast, the actual fixed point is expected to show apparent convergence under extensions of the truncation, see, e.g., \cite{Canet:2003qd}.

Quantum gravity brings an extra complication. As all metric configurations enter the path integral,  momentum scales fluctuate. Thus, how does one distinguish between UV and IR modes in order to set up the RG flow?
To this end, one introduces an auxiliary background metric and splits the expectation value of the full metric $g_{\mu\nu}$ linearly into background and fluctuation
\be
g_{\mu\nu} = \bar{g}_{\mu\nu} + h_{\mu\nu},\label{eq:linsplit}
\ee 
where the amplitude of the fluctuations is not restricted, i.e., the split is not a perturbative one. For alternative splits, see \cite{Eichhorn:2013xr,Gies:2015tca,Ohta:2016npm}. Using $\bar{g}_{\mu\nu}$, the flow equation can be set up \cite{Reuter:1996cp}: What played the role of the momentum $p^2$ in the flat background case, will now be the covariant Laplacian $-\bar{D}^2$ with respect to the metric $\bar{g}_{\mu\nu}$. Thus, UV modes are those with eigenvalues ${\rm eig} (-\bar{D}^2)>k^2$.

On the other hand, no configuration in the path integral should be distinguished -- quantum gravity should be background independent. In the present setup, shift symmetry $h_{\mu\nu}\rightarrow h_{\mu\nu}+ \gamma_{\mu\nu}$, $\bar{g}_{\mu\nu}\rightarrow \bar{g}_{\mu\nu}- \gamma_{\mu\nu}$ encodes that the effective action should only depend on one metric, namely the physical field. It is broken by  the regulator (and gauge fixing term) \footnote{In approaches to quantum gravity that focus on a ``pre-geometric" phase, where a continuum spacetime is yet to emerge from underlying discrete building blocks, the RG can be set up in a more abstract way, by coarse-graining from many to few degrees of freedom, but this typically also appears to imply the breaking of a symmetry of the model, see, e.g.,  \cite{Eichhorn:2014xaa}.}  For instance, the regulator term $h_{\mu\nu}R_k(-\bar{D}^2)^{\mu\nu\kappa\lambda}h_{\kappa\lambda}$ depends on two \emph{distinct} arguments $h_{\mu\nu}$ and $\bar{g}_{\mu\nu}$ such that they cannot be combined to a full $g_{\mu\nu}$.This leads to an enlargement of theory space. For instance, the flow of the Newton coupling read off the flow of the three-graviton vertex (i.e., by evaluating the scale-dependence of the coupling between three powers of $h_{\mu\nu}$ with two derivatives) and the background curvature term $\sqrt{\bar{g}}\bar{R}$ differ.
Crucially, the propagator on the right-hand-side of the Wetterich equation is that of the fluctuation field. 
Thus, the single-metric approximation, where the fluctuation- field- propagator is equated to the background-field propagator, introduces an artificial background dependence in the results. 
To restore background independence at the level of physical observables thus requires a careful distinction of the dependence of $\Gamma_k[\bar{g}_{\mu\nu}, h_{\mu\nu}]$ on its two distinct arguments. 
Studies in bimetric settings \cite{Manrique:2009uh}, and with a distinction of fluctuation field and background field \cite{Christiansen:2012rx} are therefore under way.

In gauge theories, the Ward identity encodes the presence of gauge symmetry. The presence of a regulator that breaks the corresponding symmetry leads to additional nontrivial terms (in addition to those from gauge fixing), e.g., \cite{Gies:2006wv}. For shift symmetry,
the modified Ward-identity encodes the difference between the fluctuation-field and background dependence of $\Gamma_k$ that is generated by the regulator and the gauge-fixing term \cite{Litim:2002ce}. Solving the flow equation and the modified Ward-identity simultaneously would restrict the flow onto a hypersurface in the enlarged theory space on which background independence arises at the level of physical observables. For studies aiming at imposing the modified Ward identity on the flow see \cite{Dietz:2015owa}.
\section{Status of asymptotic safety in pure gravity}
\label{sec:ASstatusgrav}
Before exploring asymptotic safety for quantum gravity, let us clarify a possible underlying mechanism. For  dimensionfull couplings, such as the Newton coupling, the beta functions of their dimensionless counterparts include a linear term from the canonical dimensionality, which can balance the effect of quantum fluctuations. In particular, models which are asymptotically free in their critical dimension $d_{\rm crit}$ (where the coupling is dimensionless), are asymptotically safe in 
 $d= d_{\rm crit}+ \epsilon$ ($\epsilon>0$) dimensions. For instance, the beta function for the gauge coupling $g_{\rm YM}$  in Yang-Mills theory in $d=4+\epsilon$ has the form
\be
\beta_{g_{\rm YM}} =  \frac{\epsilon}{2} \,g_{\rm YM} - \beta_0\, g_{\rm YM}^3+\mathcal{O}(g_{\rm YM}^5),\label{eq:betaYM} 
\ee
where $\beta_0$ is the one-loop coefficient. Asymptotic freedom in $d=4$ requires $\beta_0>0$, which implies that the term $\sim g_{\rm YM}$ from the canonical dimension can balance the one-loop term to induce asymptotic safety. To what value of $\epsilon$ asymptotic safety exists in Yang-Mills theory is under investigation \cite{Gies:2003ic}.

In gravity, the critical dimension is $d=2$. In $d= 2+\epsilon$, the beta function of the dimensionless Newton coupling $G = G_N\, k^{d-2}$ is similar to Eq.~\eqref{eq:betaYM},
\be
\beta_G = \epsilon\, G - \frac{38}{3}G^2+...,
\ee
see \cite{Gastmans:1977ad}. The numerical value of the quadratic term depends on the parameterization of metric fluctuations \cite{Nink:2014yya,Falls:2017cze}, but the sign does not. 
Thus, a balance of quantum fluctuations and canonical scaling could induce a scale-invariant, asymptotically safe regime in gravity.
The critical question is whether the fixed point at $G^{\ast}>0$ that exists close to $d=2$ can be extended\footnote{The fixed point in $d=4$ might of course not be smoothly connected to that in $d=2+\epsilon$, but a smooth connection to a perturbative regime provides another powerful tool to characterize the fixed point. Moreover, for a fixed point in $d=4$ that smoothly connects to one in $d=2$, the scaling exponents are given by canonical scaling dimensions plus terms which go to zero as $\epsilon \rightarrow 2$. These additional terms might be quantitatively small in $d=4$, and the fixed point could inherit a close-to-canonical scaling behavior. This is an excellent basis to set up truncations that show apparent convergence.} to $d=4$.

\begin{figure}
\begin{center}
\includegraphics[width=0.42\linewidth]{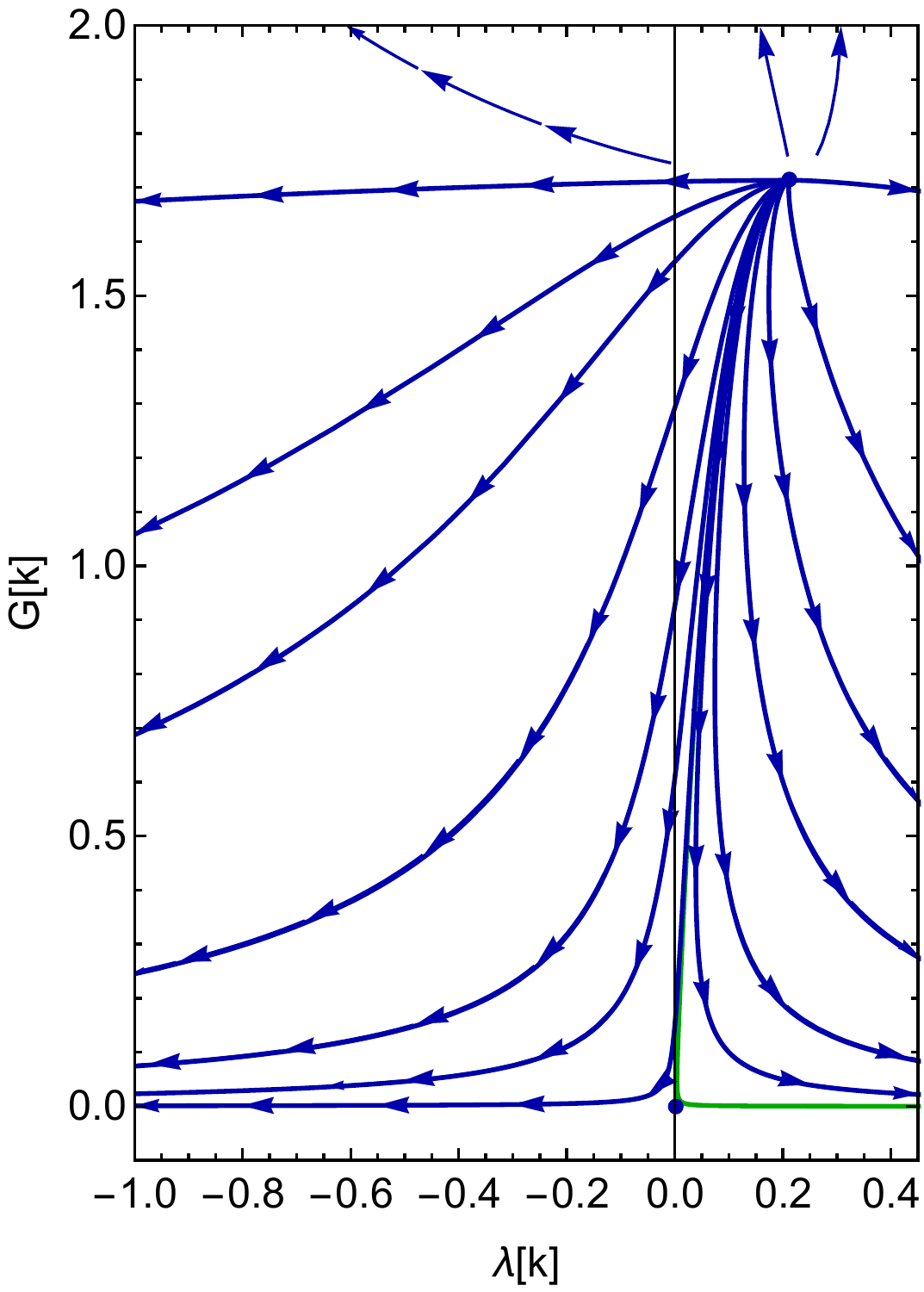}\, \quad \includegraphics[width=0.45\linewidth, clip=true, trim=11cm 8cm 12.5cm 2cm]{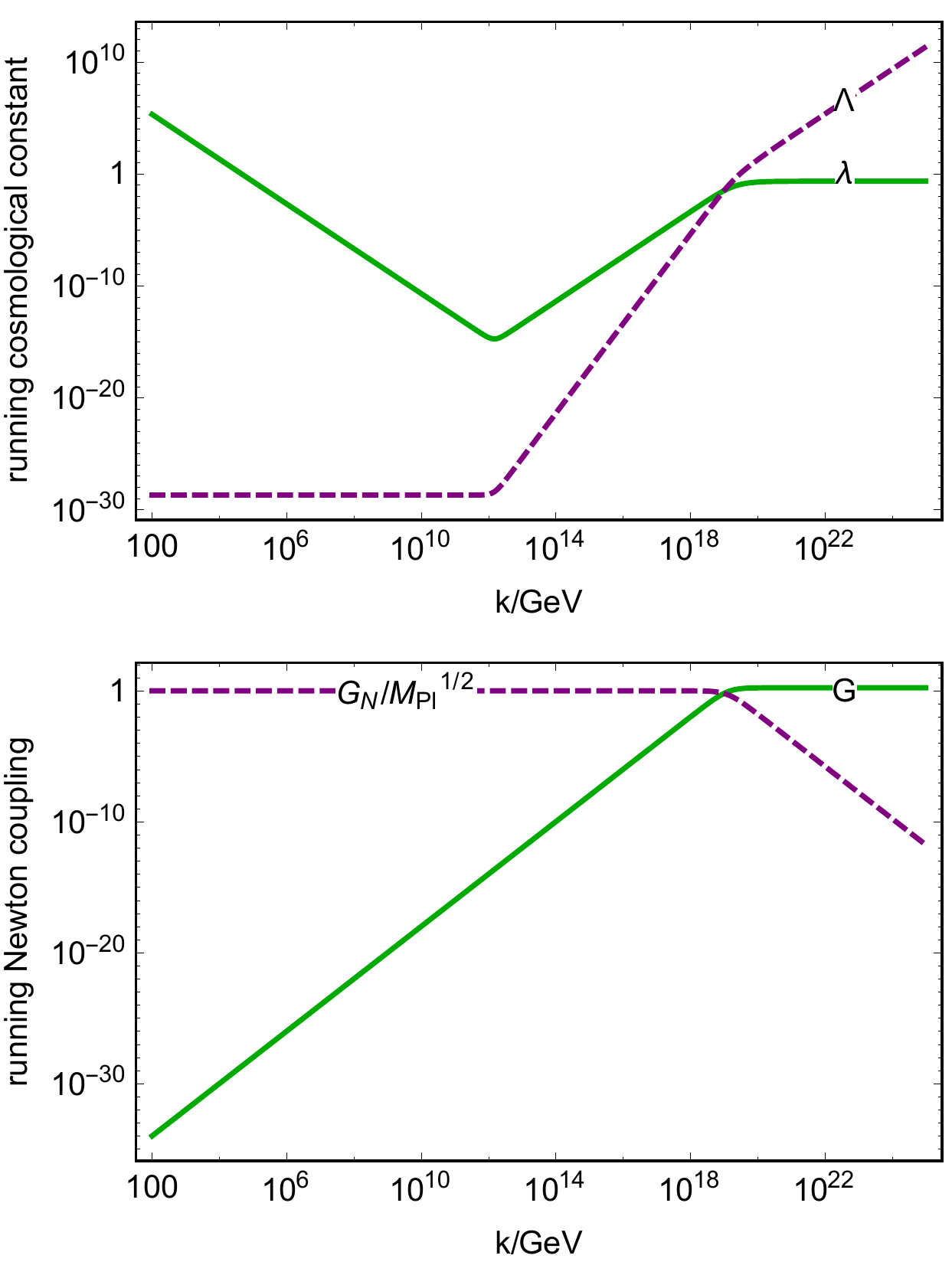}
\caption{\label{fig:flowplot} Flow of $G$ and $\lambda$ to the infrared according to eq.~\ref{eq:glambdaflow} (left panel). Arrows point towards the IR; the interacting fixed point is UV attractive in $G$ and $\lambda$. Running of the dimensionfull $G_N$ ($\Lambda$)in units of the Planck energy and the dimensionless $G$ ($\lambda$) as a function of $k/\rm GeV$ (right lower (upper) panel) along the green trajectory in the left panel.}
\end{center}
\end{figure}

Using the canonical dimension as a guide, the leading-order truncation  is  the Einstein-Hilbert action, 
\be
\Gamma_{k\, \rm EH} = -\frac{1}{16\pi\,G_N}\int d^4x\, \sqrt{g}\left(R- 2\Lambda\right) + S_{\rm gf} + S_{\rm gh},
\ee
where $G (k)= G_N(k) k^{2}$ and $\lambda(k) = \Lambda(k) k^{-2}$ are the dimensionless running couplings and $S_{\rm gf}$ is a gauge-fixing term, that is typically chosen as 

\be
S_{\rm gf}= \frac{1}{\alpha\, 32 \pi\, G_N} \int d^4x\, \sqrt{\bar{g}}\bar{g}^{\mu\nu}F_{\mu}F_{\nu}, \quad F_{\mu} =\bar{D}^{\kappa}h_{\kappa \mu} - \frac{1+\beta}{4}\bar{D}_{\mu}h,
\ee
with the corresponding Faddeev-Popov ghost term
\be
S_{\rm gh}=- \sqrt{2}\int d^4x\, \sqrt{\bar{g}}\,\bar{c}_{\mu}\left(\bar{g}^{\mu\rho}\left(\bar{D}^{\kappa}g_{\rho\nu}D_{\kappa} + \bar{D}^{\kappa}g_{\kappa \nu}D_{\rho}\right) - \frac{1+\beta}{2} \bar{D}^{\mu}D_{\nu}\right)c^{\nu}.
\ee
The beta functions for $\beta=\alpha=1$
 with a  type-Ia regulator (cf.~\cite{Codello:2008vh}) to second order in the couplings highlight the main structure, see Fig.~\ref{fig:flowplot},
\bea
\beta_G& =& 2 G - \frac{11\, G^2}{3\pi},\quad \beta_{\lambda}= - 2 \lambda + \frac{G}{2\pi} +\frac{4}{3\pi}G\, \lambda.\label{eq:glambdaflow}
\eea
The UV-attractive fixed point in $G, \lambda$  \cite{Reuter:2001ag}  exists for different choices of gauge fixing and parameterization and beyond the simple approximation in Eq.~\eqref{eq:glambdaflow}, see, e.g., \cite{Gies:2015tca}.
Extended truncations including $R^n,\, n\in \mathbb{N}$ feature a third relevant direction \cite{Lauscher:2002sq}, and near-canonical scaling for higher-order couplings \cite{Machado:2007ea,Codello:2008vh}, a posteriori supporting the choice of truncations based on the canonical dimension. The canonical dimension also appears to be a useful guide with more complicated tensor structures: adding $R^2$ and $R_{\mu\nu}R^{\mu\nu}$ to the truncation results in only one additional relevant and one additional irrelevant direction \cite{Benedetti:2009rx}. 
The presence of $R^2$ and $R_{\mu\nu}R^{\mu\nu}$ in asymptotic safety raises the question of unitarity, which requires further studies. Note that it is unlikely that a flat background, underlying conclusions on non-unitarity in four-derivative gravity \cite{Stelle:1977ry}, is the true ground state of the theory -- after all a nonvanishing cosmological constant is present in the IR, and at small scales the structure of the vacuum might be more involved \cite{Bonanno:2013dja}. Moreover, with higher-order terms in the propagator, any truncation to a finite number of derivatives leads to additional, truncation-induced zeros in the propagator \cite{Barnaby:2007ve}. Actually, the presence of \emph{infinitely} many higher-order interactions at the fixed point might be critical for unitarity, as the propagator could then feature only a pole at zero, while its expansion to finite order in powers of the momentum will always contain additional poles. 

Including the perturbative 2-loop counterterm $C_{\mu \nu\kappa \lambda}C^{\kappa \lambda \rho \sigma}C_{\rho\sigma}^{\, \, \,\, \, \mu\nu}$ with canonical dimension -4 to the Einstein-Hilbert truncation adds an irrelevant direction \cite{Gies:2016con}. Thus, in contrast to the perturbative result, at an interacting fixed point the corresponding coupling does not yield an additional free parameter.
Infinite-dimensional truncations have been explored for $f(R)$ truncations, see \cite{Benedetti:2012dx}.
Results beyond the single-metric approximation  \cite{Manrique:2009uh,Christiansen:2012rx}, which equates the fluctuation and the background propagator appear to qualitatively support the results from the single-metric approximation. The Einstein-Hilbert truncation has been explored in an ADM-setup, which naturally lends itself to a Wick-rotation to Lorentzian signature \cite{Lorentzian}.
Unimodular gravity provides a different point of view on the cosmological constant problem, and shows a fixed point  with only two relevant couplings in an expansion up to $R^{10}$ \cite{Eichhorn:2013xr}.
Imprints of asymptotically safe gravity on the structure of spacetime \cite{dimred}, the properties of black holes and cosmology \cite{cosmology} have been studied.

To summarize, results up to date might lead one to cautious optimism about the fate of gravity in a quantum-field theoretic framework: The  door for an asymptotically safe model of quantum gravity appears to be  wide open, and the step to a model of quantum gravity and matter is timely.

\section{Status of asymptotic safety for gravity and matter}
\label{sec:ASstatusgravmat}
The interplay of matter and gravity has two major aspects, namely quantum-gravity effects on matter and the impact of matter on quantum spacetime.

\subsection{Matter matters in quantum gravity}
 The analogy of the destruction of asymptotic freedom in Yang-Mills theory by quantum fluctuations of matter fields (in that case, $N_f\gtrapprox 17$ quark flavors) might suggest a similar fate of the asymptotically safe fixed point.
In fact, $N_{S/D/V}$ minimally coupled scalars, Dirac fermions and Abelian gauge fields impact the running background Newton coupling \cite{Dona:2013qba}:
\be
\beta_{G}\Big|_{\rm matter} = \frac{G^2}{6\pi}\left(N_S +2N_D -4 N_V \right),
\ee
with a  type-II cutoff \cite{Dona:2012am}, see also \cite{Eichhorn:2016vvy,Biemans:2017zca}, with a similar sign-structure for the fluctuation coupling \cite{Meibohm:2015twa,Dona:2015tnf}. This would suggest that scalars and fermions drive the Newton coupling to increasing fixed-point values, until the asymptotic-safety-inducing impact of metric fluctuations cannot compensate and the fixed point tunnels through a divergence and reappears at negative $G$. There it is presumably unphysical, as a negative UV- value of $G$ cannot be connected to positive $G$ in the IR due to the fixed point at $G=0$ which the flow cannot cross.
Including the cosmological constant leads to mass-like threshold terms in the beta functions. Depending on the sign of the ``mass", gravity fluctuations can be  enhanced or suppressed.  Here, the picture is not as clear-cut, as background-results and fluctuation field results appear to differ in certain aspects \cite{Meibohm:2015twa}. \\
In all studies to date, minimally coupled Standard Model fields admit an asymptotically safe fixed point for gravity. Thus asymptotically safe quantum gravity might pass this important observational consistency test and persist under the impact of the observed matter degrees of freedom. 

\subsection{Gravity- induced UV completion of the Standard Model}

\begin{figure*}[!t]
\begin{center}
\includegraphics[width=0.9\linewidth,clip=true,trim=3cm 14.5cm 0cm 8.2cm]{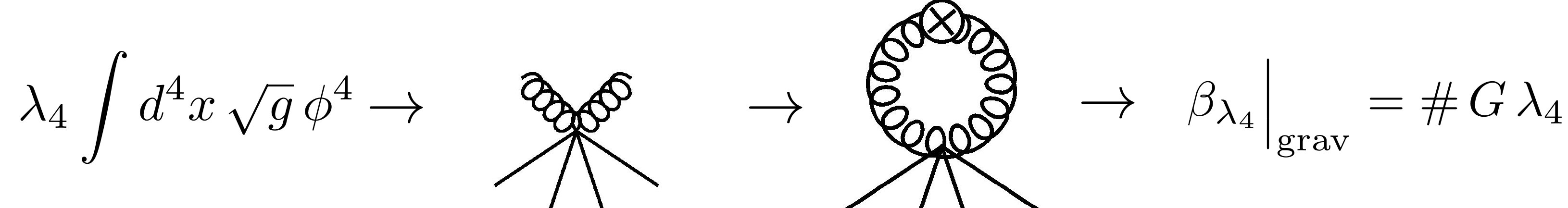}
\caption{\label{fig:betalambda4} The factor $\sqrt{g}$ provides a two-graviton-4-scalar vertex $\sim \lambda_4$, yielding a tadpole contribution $\sim \lambda_4\, G$ to $\beta_{\lambda_4}$ on the rhs of the Wetterich equation.}
\end{center}
\end{figure*}

There are intriguing hints that the answer to "Can asymptotically safe quantum gravity trigger a predictive, observationally viable UV completion for the Standard Model?"  might be affirmative.
These hints arise from the presumed leading-order effects of quantum gravity on matter beta functions. 
Extending truncations beyond these is mostly a task for the future.
The crucial gravitational term acts like a scaling dimension, i.e., it is \emph{linear} in the matter coupling.  Thus it can balance the one-loop matter terms, which involve higher powers of the matter coupling, generating an interacting fixed point.
The linear gravity contribution differs for diverse matter interactions, but is generically present. The simplest example is $\lambda_4\, \phi^4$, where $\sqrt{g}$ provides a four-scalar-two-graviton vertex $\sim \lambda_4$, sourcing a tadpole contribution $\sim G\, \lambda_4$, as each metric propagator is $\sim G$, cf.~Fig.~\ref{fig:betalambda4}. 

Specifically, for a canonically marginal matter coupling $g_i$
\be
\beta_{g_i}= \#_{\rm grav}\, G\, g_i + \#_{\rm matter} g_i^{\#_i}+...\label{eq:betamatter_schematic}
\ee
with $\#_i=2$ (scalar quartic coupling) or $\#_i=3$ (Yukawa, gauge couplings). \\

\begin{figure}[!t]
\begin{center}
\includegraphics[width=0.32\linewidth]{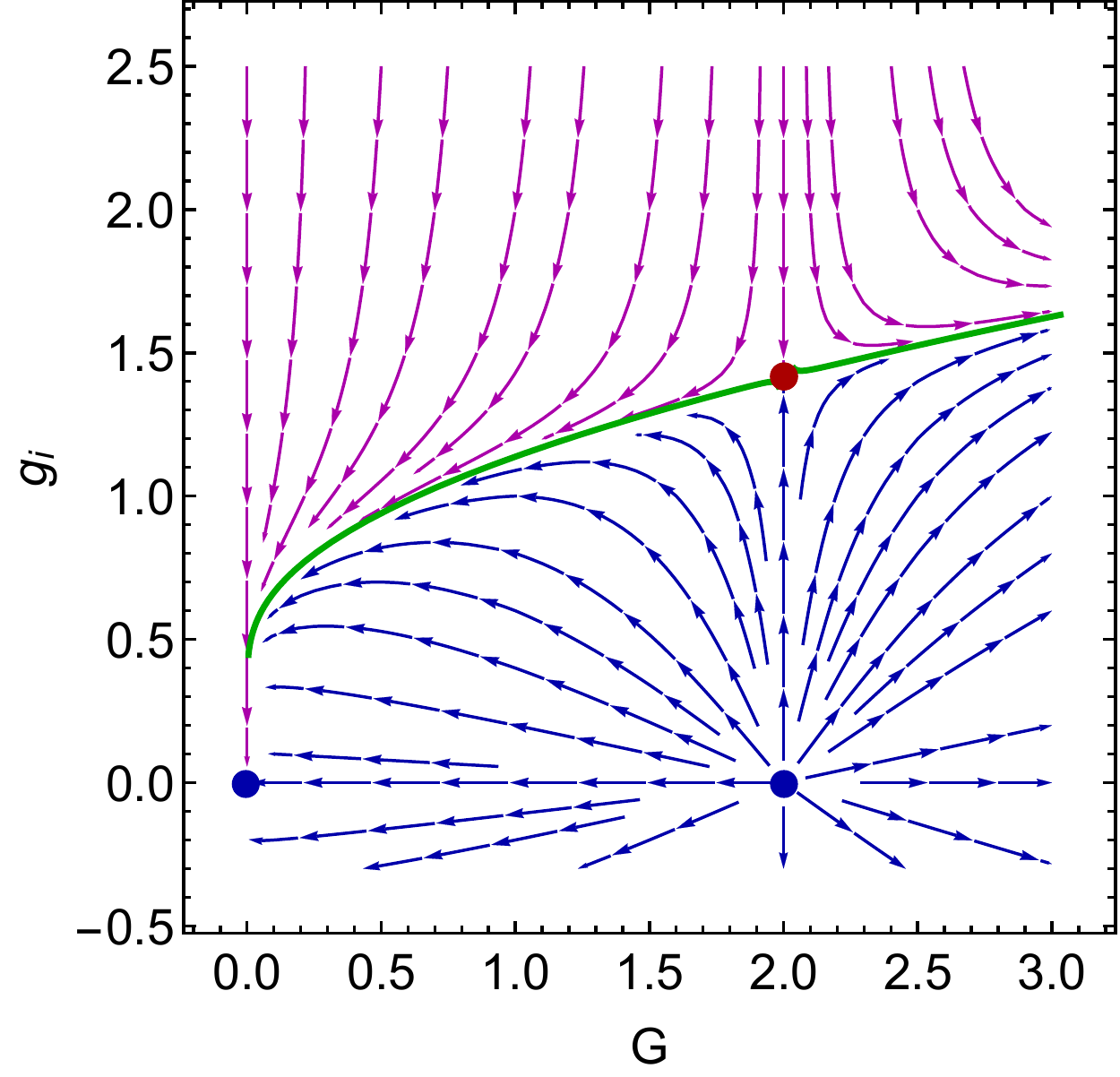} \quad \includegraphics[width=0.45\linewidth]{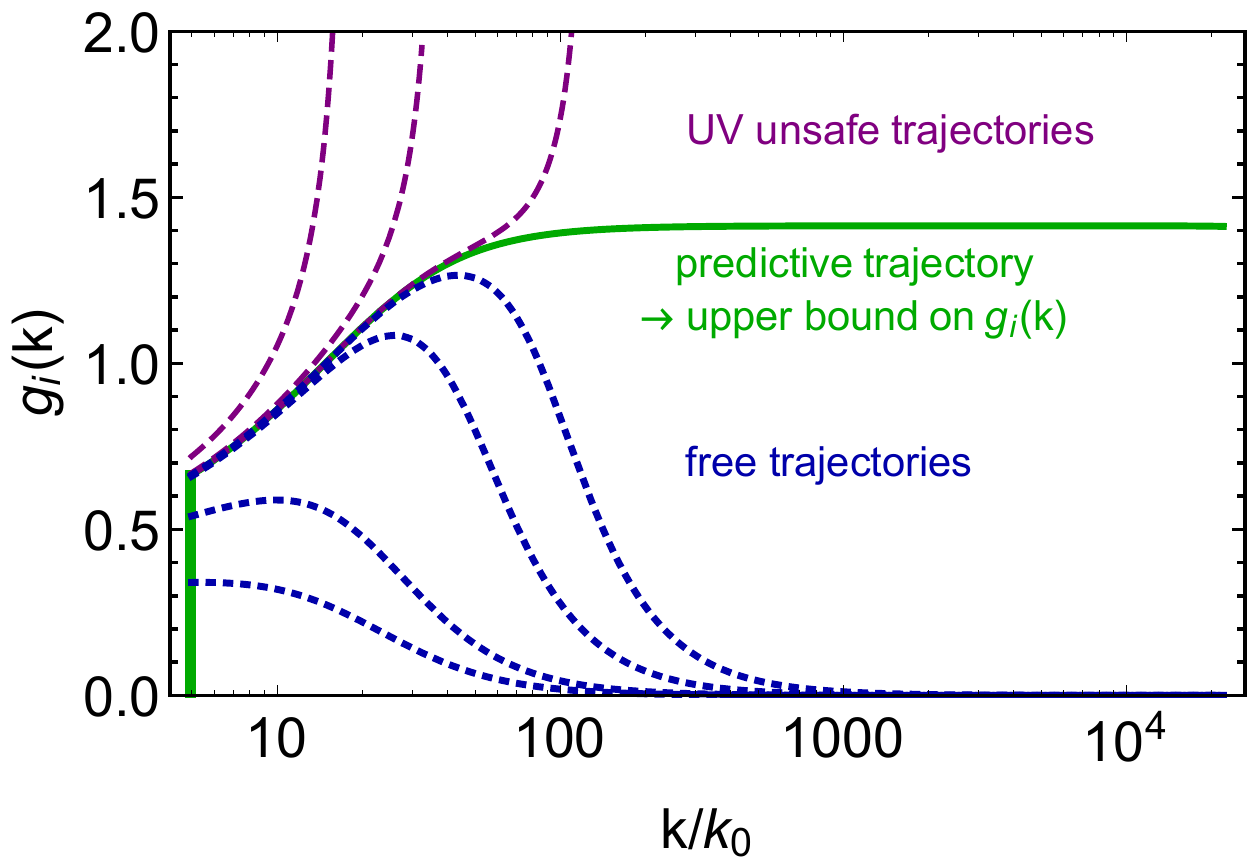}
\caption{\label{fig:upperbound_illustration} Left panel: flow of Newton coupling $G$ and matter coupling $g_i$ for case 2: The flow to the IR for  $\beta_G = 2G-G^2$, $\beta_{g_i}=g_i^3- G\, g_i$ features UV-complete trajectories below the thick green one. None of the purple trajectories above the thick green line are UV complete, as the repulsive nature of the fully interacting fixed point at $G=2, g_i = \sqrt{2}$ shields them from the UV-complete regime where they could reach the free fixed point. Along the thick green trajectory, $g_i$ is fixed as a function of $G$. Right panel: Trajectories with an IR value $G(k/k_0=5)=0.1$: There is one unique IR value of $g_i = g_{i\, \rm crit}$, for which the interacting fixed point can be reached (green trajectory), whereas IR values above that upper bound hit a divergence at a finite scale. IR values below the lower bound (indicated by the green vertical line) can be reached from an asymptotically free fixed point.}
\end{center}
\end{figure}

The coupling is asymptotically free in the absence of gravity if $\#_{\rm matter}<0$. Quantum gravity supports asymptotic freedom if $\#_{\rm grav}<0$ (case 3 in Tab.~\ref{tab:MFP}), and the IR value of the coupling is a free parameter.\\
Conversely, quantum gravity destroys asymptotic freedom if $\#_{\rm grav}>0$ (case 4 in Tab.~\ref{tab:MFP}). In that case, Eq.~\eqref{eq:betamatter_schematic} contains a UV-attractive interacting fixed point, i.e., the IR value of the coupling is a free parameter. Both signs for $\#_{\rm grav}$ thus lead to phenomenologically viable fixed points if $\#_{\rm matter}<0$.\\
\begin{table}[!b]
\begin{tabular}{c|c|c|c|c|c|c}
case &${\rm sgn}(\#_{\rm matter})$& ${\rm sgn}(\#_{\rm grav})$ &free FP &free FP & int.~FP& int.~FP\\
& &  & (free par.) & (predictive) & (free par.) & (predictive)\\\hline\hline
1&+ & + & X & \checkmark & X & X \\ \hline
2& + & - & \checkmark & X &X & \checkmark \\ \hline
3&- & - & \checkmark & X & X & X \\ \hline
4&- & + & X & \checkmark & \checkmark & X \\ \hline\hline
\end{tabular}
\caption{Existence of free/interacting fixed point and predictivity/existence of a free parameter for the matter coupling for different cases in Eq.~\eqref{eq:betamatter_schematic}, see also illustration in Fig.~\ref{fig:illcases}.
For case 2, the fixed-point structure implies an upper bound on the IR value of the coupling.\label{tab:MFP}}
\end{table}
\begin{figure*}[!t]
\includegraphics[width=0.45\linewidth]{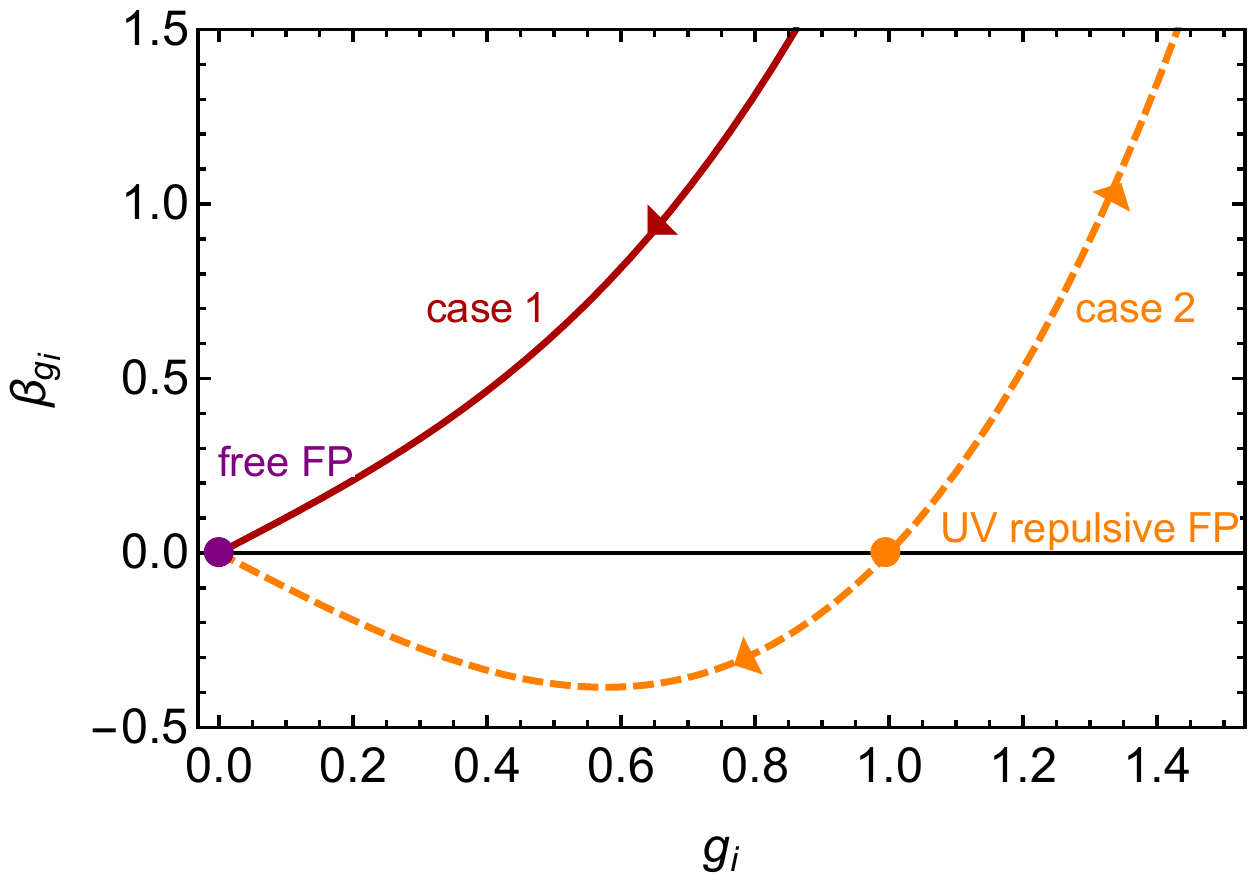}\quad\includegraphics[width=0.45\linewidth]{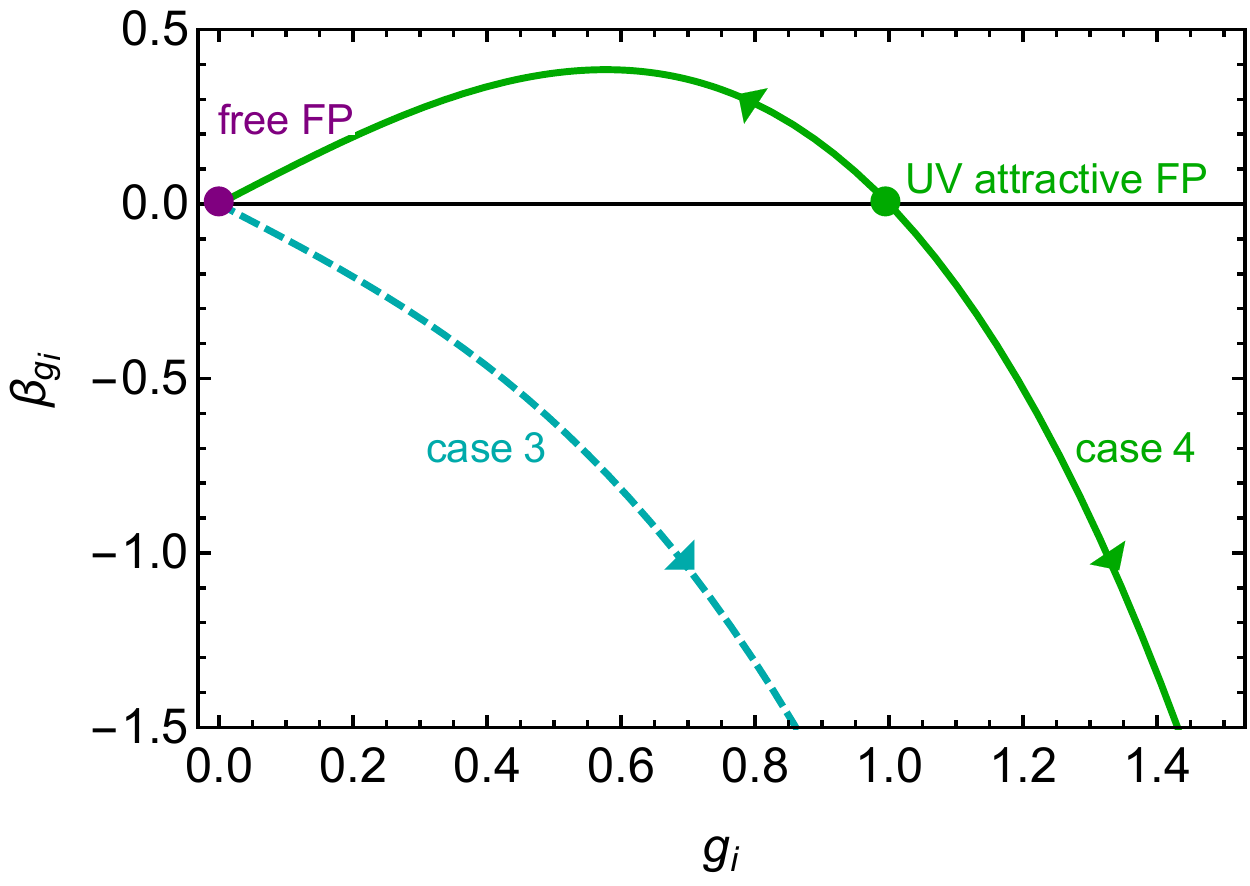}
\caption{\label{fig:illcases}Illustration of cases 1 to 4 in Tab.~\ref{tab:MFP}. Case 2 leads to an upper bound on the IR value of the coupling, see Fig.~\ref{fig:upperbound_illustration}.}
\end{figure*}
The situation differs for  $\#_{\rm matter}>0$. At one loop, the coupling grows towards the UV. For $\#_{\rm grav}>0$, the coupling is still irrelevant at the free fixed point, (case 1 in Tab.~\ref{tab:MFP}). Thus, on the only UV complete RG trajectory the coupling is zero as long as quantum fluctuations of gravity are present. This yields the prediction $g_i(k \approx M_{\rm Planck})=0$. For $k< M_{\rm Planck}$, it can be driven away from zero by additional matter contributions in Eq.~\eqref{eq:betamatter_schematic}.
Actually, the Higgs quartic coupling (nearly) fulfills this condition \cite{Bezrukov:2012sa,Buttazzo:2013uya}.  Thus asymptotic safety might predict a Higgs mass close to the observed value \cite{Shaposhnikov:2009pv}. \\
On the other hand, if $\#_{\rm grav}<0$, quantum gravity induces asymptotic freedom, i.e., the free fixed point is UV attractive, and can be reached from a range of low-energy values of the coupling (case 2 in Tab.~\ref{tab:MFP}). Simultaneously, quantum fluctuations of matter and gravity balance at an interacting fixed point at $g_i^{\ast}>0$. This fixed point entails an upper bound on the low-energy value of the coupling, \cite{Eichhorn:2017ylw,Versteegen}: The interacting fixed point is UV repulsive, i.e., it can only be reached from a unique IR- value of the coupling $g_i=g_{i\, \rm crit}$. All IR values below  $g_{i\, \rm crit}$ can be reached from the free fixed point in the UV. None of the IR values above $g_{i\, \rm crit}$ can be reached from a UV complete model, cf.~Fig.~\ref{fig:upperbound_illustration}: For these IR values, matter fluctuations are too strong, and cannot be counteracted by  quantum-gravity fluctuations. Thus no fixed point can be reached in the UV. The interacting fixed point therefore shields all IR values above the upper bound, $g_i (k_{\rm IR})>g_{i\, \rm crit}$, from the UV-complete regime.

Note that due to the nonzero dimension of the Newton coupling the gravity-contribution to the beta functions is gauge-dependent and non-universal; for the corresponding discussion in the distinct case of perturbation theory, see \cite{EFTgauge}. Gauge-independent observables can arise despite gauge dependence of beta functions \cite{Antoniadis:1985ub}.
 In particular, gauge-dependence in gravity-contributions to matter beta functions might cancel against a corresponding gauge dependence in the gravitational fixed-point values, guaranteeing gauge independence of IR observables.
 As beta functions are not physical observables, their gauge-/scheme-dependence is not unexpected -- in fact, in the Standard Model without gravity, scheme dependence sets in at three loops. 
Gauge dependence even provides tests of the quality of a truncation, see, e.g., \cite{Gies:2015tca}. 
 
It is intriguing that a UV complete gravity-matter model features a gravity contribution for the Standard Model couplings that acts similarly to dimensional reduction: Just like a scaling dimension, the gravity contribution is linear in the matter coupling with a sign as it would be for $d<4$. 
One might speculate that a UV complete gravity-matter model could hinge on a dynamical mechanism that moves the system towards the critical dimension of gravity, $d=2$, leaving imprints in beta functions and the spectral dimension \cite{Carlip:2016qrb,dimred}.

\subsubsection{Quantum-gravity effects on gauge couplings}

Abelian gauge couplings presumably exhibit the triviality problem \cite{GellMann:1954fq}, indicating the need for new physics. In the Abelian sector of the Standard Model, the corresponding breakdown is expected beyond the Planck scale, suggesting that the missing physics could be quantum gravity. 
 \begin{figure*}[!t]
 \begin{minipage}{0.5\linewidth}
 \includegraphics[width=0.8\linewidth]{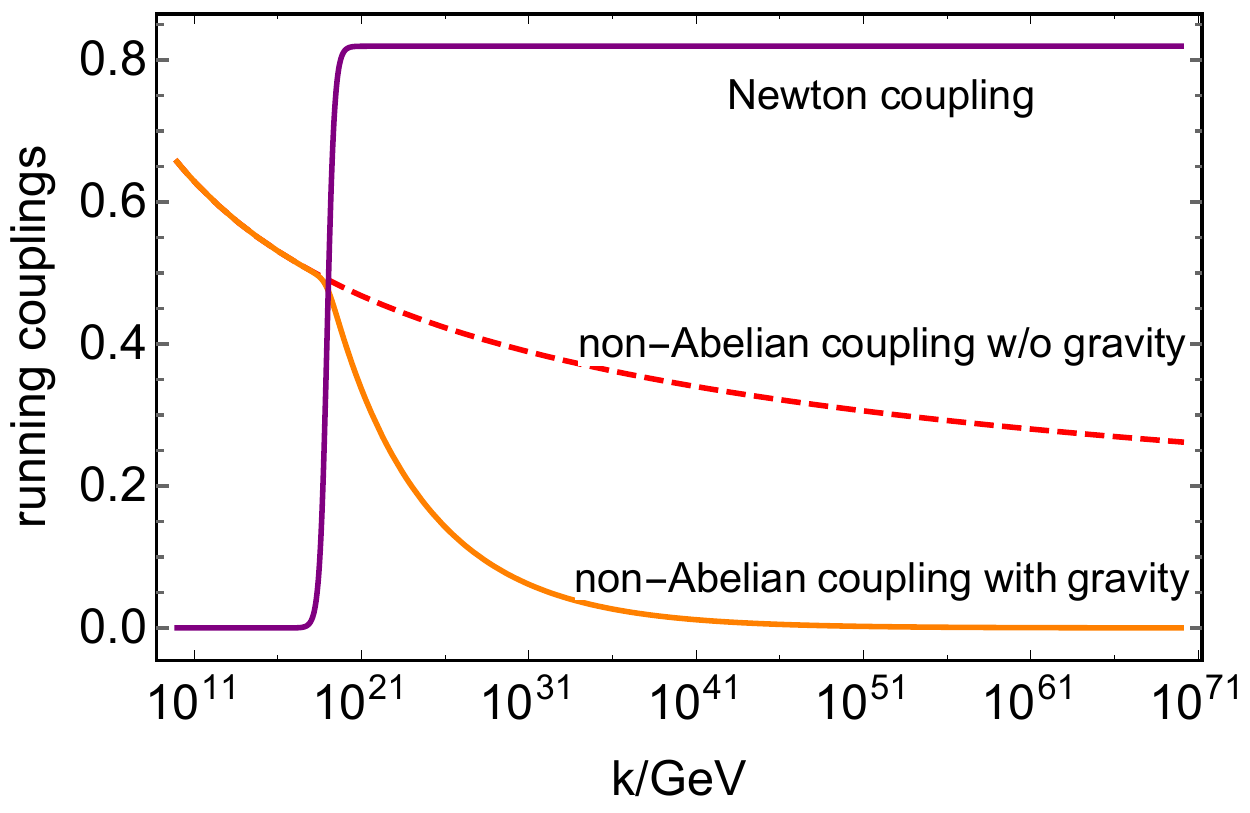}
 \end{minipage}
 \begin{minipage}{0.4\linewidth}
 \caption{\label{fig:gaugecouplings} Running non-Abelian gauge coupling without/with gravity (red dashed/ orange continuous line).}
 \end{minipage}
 \end{figure*}
Indeed, for (non-)Abelian gauge fields, results in truncations including the Einstein-Hilbert term suggest $\#_{\rm grav}<0$ \cite{Harst:2011zx,Daum:2009dn,Folkerts:2011jz,Christiansen:2017gtg,Versteegen}. Thus, asymptotic freedom is induced in gauge couplings in this approximation, hinting at a  solution to the triviality problem. In the Abelian case, an interacting fixed point appears, leading to an upper bound on the IR value of the gauge coupling in the vicinity of the observed value for appropriate values of $G, \Lambda$ \cite{Versteegen}, cf.~illustration in Fig.~\ref{fig:upperbound_illustration}.
 For non-Abelian gauge couplings, the approach to asymptotic freedom accelerates beyond the Planck scale, as the logarithmic running is turned into a power-law running with exponent $\theta =-\#_{\rm grav}G$ by quantum-gravity effects, cf.~Fig.~\ref{fig:gaugecouplings}.

 \subsubsection{Quantum-gravity effects on the Higgs-Yukawa sector}
For the Yukawa couplings the sign of the presumed leading contribution depends on the gravitational fixed-point values \cite{Zanusso:2009bs,Eichhorn:2016esv,Eichhorn:2017eht,Eichhorn:2017ylw}, cf.~Fig.~\ref{fig:wgb}. For values with $\#_{\rm grav}>0$, there is a UV repulsive free fixed point and thus -- unless there are effects from curvature of the critical hypersurface -- the Yukawa coupling is forced to zero at the Planck scale, resulting in vanishing fermion masses. \\
 In the Einstein-Hilbert truncation, 
 with  $\lambda<\lambda_{\rm crit}<0$, $\#_{\rm grav}<0$ holds \cite{Eichhorn:2017ylw}, realizing the scenario in Fig.~\ref{fig:upperbound_illustration}. In the single-metric approximation, UV complete flows for gravity and the Standard Model starting from the finite fixed-point value of the top Yukawa and Higgs quartic coupling lead to unique IR values for top and Higgs mass. As quantum gravity does not break chiral symmetry, fermion masses to remain small compared to the Planck scale \cite{Eichhorn:2011pc,Eichhorn:2016esv}, and the top mass comes out close to the experimentally observed value \cite{Eichhorn:2017ylw}.

\subsection{Where are the interactions at the asymptotically safe fixed point?}
Asymptotically safe gravity is fully interacting in the UV. One might expect that interactions percolate into the matter sector. However, several of the Standard Model couplings become asymptotically free under the impact of quantum gravity. So where are the interactions hiding? The answer hinges on global symmetries \cite{Eichhorn:2017eht}: In truncations, quantum gravity  indeed induces all possible interactions, but the possibilities are determined by symmetry: Interactions respecting the global symmetry of the kinetic terms do typically \emph{not} have a free fixed point \cite{Christiansen:2017gtg,Eichhorn:2016esv,Eichhorn:2017eht,Eichhorn:2011pc,Eichhorn:2012va}. However, interactions which break the global symmetry of the kinetic terms, e.g., all marginal interactions in the Standard Model, feature a free fixed point, and might additionally have an interacting fixed point, such as, e.g., for the Yukawa coupling. The mechanism is simple: Starting from the kinetic term for a matter field, one-loop diagrams with $2n$ external matter fields and internal gravitons can be built,  yielding a contribution to the flow of a $2n$-matter interaction which is \emph{independent} of the corresponding coupling itself. Thus, setting the coupling to zero does not yield a fixed point. Instead, the gravity contributions shift the free fixed point to an interacting one \cite{Eichhorn:2011pc,Eichhorn:2012va}. Hence, if quantum scale invariance in the UV can be realized, it appears to necessitate the presence of specific matter selfinteractions.  These are typically momentum-dependent and thus higher order from a perturbative point of view.

\begin{figure*}[!t]
\begin{center}
\includegraphics[width=0.4\linewidth]{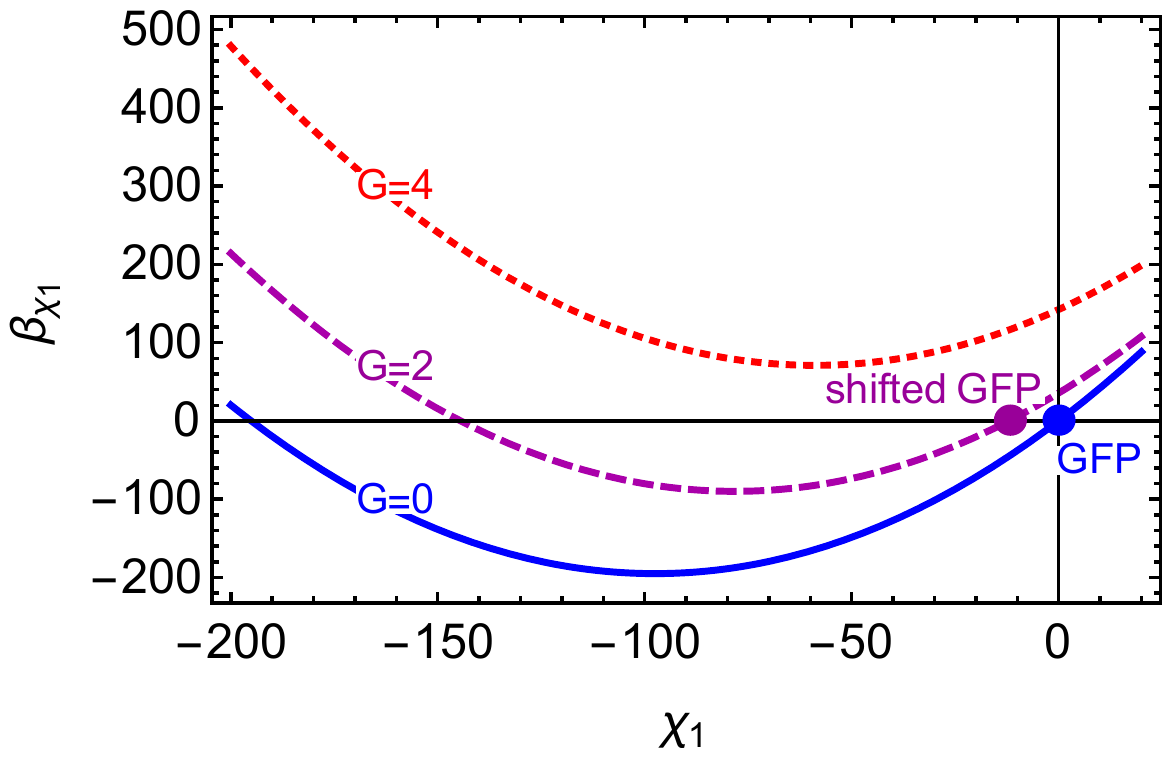}\quad \includegraphics[width=0.38\linewidth]{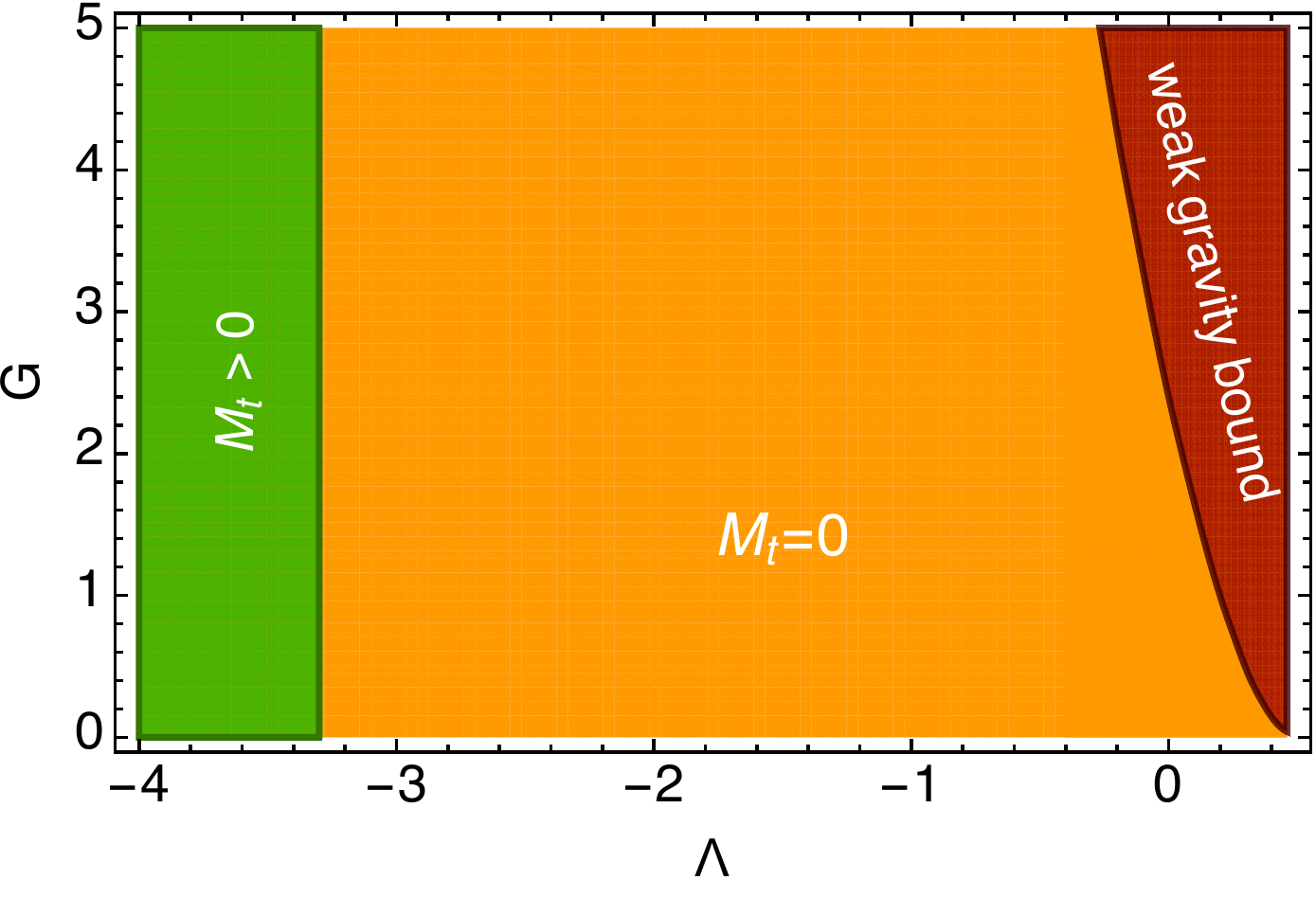}
\end{center}
\caption{\label{fig:wgb} Left: Quartic fermion-scalar interactions parameterized by $\chi_1$ (see \cite{Eichhorn:2016esv,Eichhorn:2017eht}) feature a shifted free fixed point at $G_{\rm crit}>G>0$ and no real fixed point beyond $G_{\rm crit}\approx 3.2$. Right: In the $G-\Lambda$ plane, the bound on the effective gravitational strength also depends on  $\Lambda$. Regimes with a finite (green) and vanishing (orange) top mass are shown (see \cite{Eichhorn:2017ylw}).}
\end{figure*}

However, the reality of the fixed-point values for those couplings is not guaranteed. Thus a study of these higher-order interactions is critical to reveal whether a matter-gravity fixed point can indeed exist. For instance, for quartic matter couplings the beta function is a parabola in the coupling, and the gravity contribution introduces a shift. Depending on its sign, that shift can destroy the fixed point, if quantum gravity fluctuations become too strong, cf.~Fig.~\ref{fig:wgb}, see \cite{Eichhorn:2016esv,Christiansen:2017gtg,Eichhorn:2017eht}. These results suggest that the gravitational fixed-point values have to satisfy the weak gravity bound, cf.~right panel in Fig.~\ref{fig:wgb}. There, threshold effects are included: Negative $\Lambda$ acts like a positive mass and suppresses metric fluctuations, weakening the effective gravitational strength. Within truncations, the weak-gravity bound appears to be satisfied, see \cite{Eichhorn:2017eht}.

\section{Outlook}
Based on results in simple gravity-matter truncations, the existence of a highly predictive asymptotically safe gravity-Standard-Model-fixed point appears potentially possible.  It is critical to investigate the system at higher orders in the truncation to confirm the existence and properties of the fixed point. Further, it is crucial to reach quantitative precision in the prediction of low-energy observables in the matter sector. This could trigger critical progress on quantum gravity: If the quantitative predictions for low-energy observables  converge away from the experimental values, the model containing only gravity and the Standard Model is ruled out. Then, asymptotic safety would only remain viable if one could show that new physics that is observationally viable can evade theoretical constraints such as the weak-gravity bound -- assuming that it persists in the extended setting -- while at the same time featuring an asymptotically safe fixed point and altering the low-energy values of all predictions such that these match the observed values. Contrary to the often-heard claim that experimental tests of quantum gravity are impossible, or only possible in special cases, e.g., for strong violations of Lorentz symmetry, demanding compatibility of a microscopic model with all low-energy observations, in particular those in the matter sector, appears to emerge as a potentially strong test of quantum gravity.  As physics is and will remain an experimental science -- even when it comes to quantum gravity! -- and theoretical progress is tied to observational tests, this is cause for optimism.

\begin{acknowledgements}
 I  thank the organizers of the workshop on Black Holes, Gravitational Waves and Spacetime Singularities for the invitation to a particularly inspiring workshop.
It is a pleasure to thank N.~Christiansen, P.~Don\`a, H.~Gies, A.~Held, P.~Labus, S.~Lippoldt, J.~Pawlowski, R.~Percacci, M.~Reichert and F.~Versteegen for enjoyable and fruitful collaborations on gravity-matter systems, some part of which is reflected in these notes. I am indebted to A.~Held and F.~Versteegen for help in making this summary (hopefully) more understandable.
I acknowledge funding by the DFG within the Emmy-Noether-program under grant no.~Ei-1037-1 and support by the Perimeter Institute for Theoretical Physics through the Emmy-Noether-visiting fellow program.
\end{acknowledgements}

\end{document}